\documentclass[10pt,journal,compsoc]{IEEEtran}
\usepackage[table]{xcolor}

\usepackage{booktabs}      
\usepackage{array}
\usepackage{colortbl}
\usepackage{multirow}
\usepackage{multicol}
\usepackage{graphicx}
\usepackage{subfigure}

\usepackage{amsmath,amsthm,amsfonts,amssymb}
\usepackage{mathrsfs}
\usepackage{bm}
\usepackage{bbm}
\usepackage{algorithmic}
\usepackage{float}

\usepackage{caption}
\usepackage{accents}
\usepackage{diagbox}
\usepackage{wrapfig}
\usepackage{comment}

\usepackage{pifont,bbding,makecell}

\usepackage{soul}

\theoremstyle{plain}

\theoremstyle{definition}

\theoremstyle{remark}

\definecolor{darkred}{rgb}{0.6,0.0,0.0}
\definecolor{darkgreen}{rgb}{0,0.50,0}
\definecolor{lightblue}{rgb}{0.0,0.42,0.91}
\definecolor{orange}{rgb}{0.99,0.48,0.13}
\definecolor{grass}{rgb}{0.18,0.80,0.18}
\definecolor{pink}{rgb}{0.97,0.15,0.45}

\definecolor{cite_color}{rgb}{0,0.50,0}
\definecolor{link_color}{RGB}{190,0,60} 
\definecolor{url_color}{RGB}{153,102,0}
\definecolor{emp_color}{RGB}{0,0,255}
\definecolor{ref_color}{RGB}{0,0,255}

\usepackage[vlined,boxed,commentsnumbered,ruled,linesnumbered]{algorithm2e}
\usepackage{xcolor}

\ifCLASSOPTIONcompsoc
  \usepackage[nocompress]{cite}
\else
  \usepackage{cite}
\fi
\ifCLASSINFOpdf
\else
\fi

\begin{document}

\title{
 Towards Practical Large-scale Dynamical Heterogeneous Graph Embedding: Cold-start Resilient Recommendation
\author{Mabiao~Long$^\ast$,~Jiaxi~Liu$^\ast$,~Yufeng~Li,~Hao~Xiong$^\dagger$,~Junchi~Yan$^\dagger$~\IEEEmembership{Senior Member,~IEEE},~Kefan~Wang,~Yi~Cao,~and~Jiandong~Ding.
\thanks{M. Long, J. Liu, Y. Li and J. Yan are with Department of Computer Science and Engineering, MoE Key Lab of Artificial Intelligence, AI Institute, Shanghai Jiao Tong University, Shanghai, China. J. Yan is also affiliated with Zhejiang Lab, Hangzhou, China. H. Xiong is with Artificial Intelligence Innovation and Incubation Institute, Fudan University, Shanghai, China. K. Wang, Y. Cao, and J. Ding joined this project as independent researchers not affiliated with any institution.\protect
} 
\thanks{$^\ast$ denotes equal contribution.}
\thanks{$^\dagger$: H. Xiong and J. Yan are co-corresponding authors.}
\thanks{E-mail: yanjunchi@sjtu.edu.cn, haoxiong@fudan.edu.cn}
\thanks{Preprint.}
}}

\markboth{Journal of \LaTeX\ Class Files,~Vol.~14, No.~8, December~2025}%
{Shell \MakeLowercase{\textit{et al.}}: Bare Demo of IEEEtran.cls for IEEE Journals}

\IEEEtitleabstractindextext{
\begin{abstract}
Deploying dynamic heterogeneous graph embeddings in production faces key challenges of scalability, data freshness, and cold-start. This paper introduces a practical, two-stage solution that balances deep graph representation with low-latency incremental updates. Our framework combines HetSGFormer, a scalable graph transformer for static learning, with Incremental Locally Linear Embedding (ILLE), a lightweight, CPU-based algorithm for real-time updates. HetSGFormer captures global structure with linear scalability, while ILLE provides rapid, targeted updates to incorporate new data, thus avoiding costly full retraining. This dual approach is cold-start resilient, leveraging the graph to create meaningful embeddings from sparse data. On billion-scale graphs, A/B tests show HetSGFormer achieved up to a 6.11\% lift in Advertiser Value over previous methods, while the ILLE module added another 3.22\% lift and improved embedding refresh timeliness by 83.2\%. Our work provides a validated framework for deploying dynamic graph learning in production environments.

\end{abstract}

}
\maketitle


%
\IEEEpeerreviewmaketitle

\begin{figure}[!tb]
    \centering
    \includegraphics[width=0.45\textwidth]{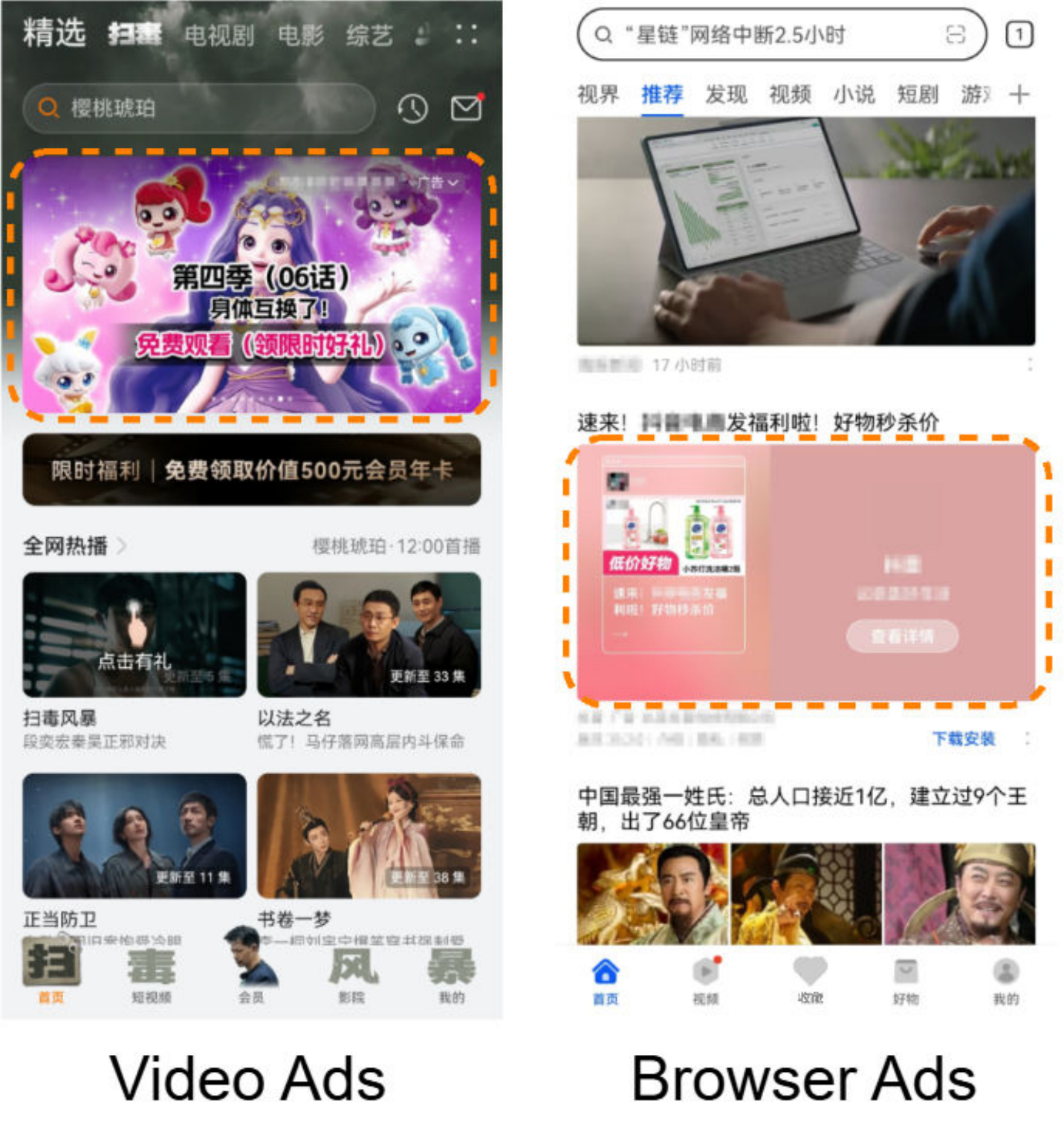}
    \caption{The examples of diverse advertisement formats (highlighted in \textcolor{orange}{orange dashed box}) in apps on a smartphone.}
    \label{fig:ads}
\end{figure}

\IEEEraisesectionheading{\section{Introduction}}\label{sec:intro}
\IEEEPARstart{M}odern smartphones serve a consumer business with billions of active users and are equipped with an ecosystem of first-party applications, browsers, news feeds, and more. The ecosystem offers diverse advertising formats (with concrete examples shown in Fig. \ref{fig:ads}), including lock screens, home screens, browsers, and incentive videos, forming a large-scale, complex, and dynamic recommendation network. This network serves as a platform that connects media traffic across the entire ecosystem, delivering a premium marketing experience to users while unlocking significant commercial value. However, unlike global ad platforms that track users across numerous third-party sites, a device vendor must operate within a much smaller data envelope: only in-device signals (app installs, click logs, and coarse-grained demographics) are legally and technically accessible. This scarcity makes high-precision targeting exceptionally challenging, particularly for new users (cold start) and in the face of rapidly shifting interests (temporal drift). Consequently, the advertising stack must extract maximum predictive power from sparse, high-volatility data while respecting strict privacy and latency constraints, an engineering challenge that motivates the present project.

Graph embedding, one of the key techniques in modern recommendation systems, converts rich, heterogeneous interactions among users, items, and their associated side information into low-dimensional, dense vectors that preserve node proximity \cite{sha2021commercial, grad2017graph}. By learning representations that encode collaborative signals (e.g., co-occurrence, co-purchase, or shared context) as well as content and temporal patterns, graph embeddings compress sparse, high-cardinality relational data into a form that downstream ranking and retrieval models can exploit for recommendation \cite{deng2022recommender}. Dynamical heterogeneous graph embedding (DHGE) \cite{yang2020dyhan,wang2022survey,ji2021dynamic}, designed for systems where both the graph topology and node attributes evolve over time, extends static heterogeneous graph embedding by jointly modeling (i) multi-typed entities (e.g. users, ads, apps, contexts), (ii) multi-typed relations (e.g. install, click, view, co-occurrence), and (iii) their continuous drift. Instead of retraining from scratch at each time step, DHGE incrementally updates latent vectors via an efficient time-aware fusion mechanism that blends new events with historical summaries.

Typically, a DHGE pipeline \cite{wang2022survey,ji2021dynamic} has two stages:
\textbf{i) Static Graph Learning (SGL)} – a heavy, full-graph optimization that produces high-quality seed embeddings for every node in the graph.
\textbf{ii) Incremental Graph Learning (IGL)} – a lightweight streaming update that revises only the embeddings affected by new events or temporal drift.
SGL is indispensable as it seeds the latent space with global structure, but its computational footprint makes it too slow for real-time use. IGL, while fast, is inherently local and gradually drifts if run indefinitely without global re-anchoring. We therefore alternate the two phases: a full SGL sweep is executed at coarse intervals to recalibrate the entire space, while the cheaper IGL steps are interleaved densely to propagate fresh signals within milliseconds of arrival.

\begin{figure}[!tb]
    \centering
    \includegraphics[width=0.5\textwidth]{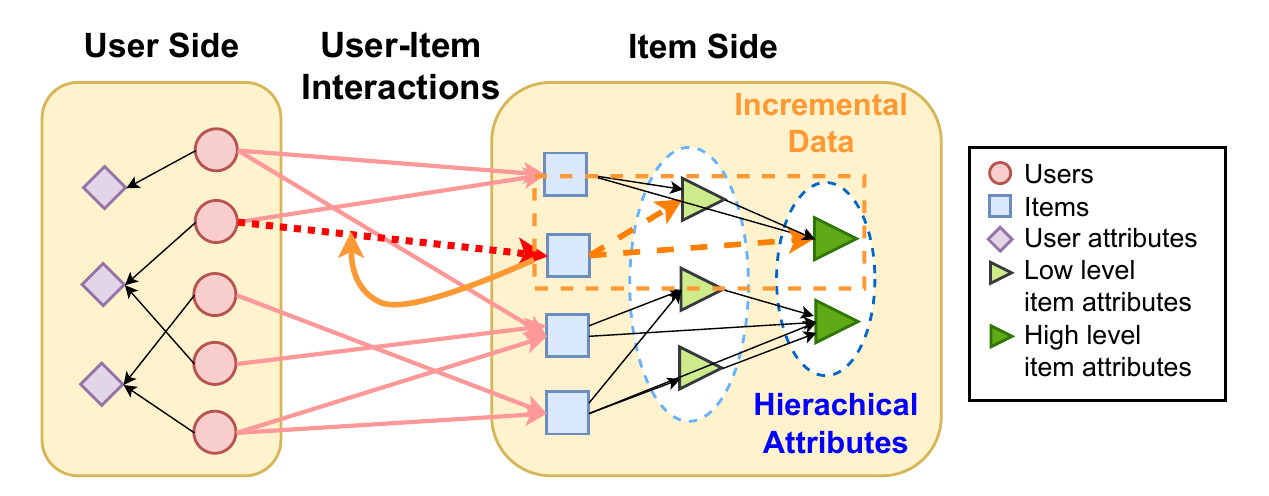}
    \caption{An illustrative toy heterogeneous recommendation graph of our data. Users and ads appear as distinct node types, each with its own feature hierarchy: low-level signals feed upward into more abstract concepts, shown as stacked layers. The \textcolor{orange}{dashed orange box} highlights the incremental data; the \textcolor{red}{dashed red edge} signals the need for a fast, cold-start mechanism that can yield embeddings for incremental nodes on the fly.}
    \label{fig:heterograph}
\end{figure}

We distill the key challenges encountered in designing and deploying the two-stage DHGE algorithm in the real-world large-scale data environments from the following three fine-grained aspects:

\textbf{i) Cold-start problem.}  
It arises from two complementary gaps. On the one hand, \textbf{attribute scarcity}: a large fraction of user or item nodes arrive with incomplete or entirely missing attributes (e.g., age, gender, declared interests for users or selling points, luxury, different levels of categories for items), obliging the model to learn coherent embeddings for both richly-attributed and attribute-void vertices within a single architecture. On the other hand, \textbf{behavior scarcity}: newly registered users typically exhibit only a handful of ad impressions, clicks, or installs, and newly added items often receive minimal engagement (as shown in Fig. \ref{fig:heterograph}), yielding extremely sparse historical sequences that challenge the model’s ability to infer accurate, personalized representations.

\textbf{ii) Efficiency and scalability.}  
Our production graph encompasses billions of user nodes and several tens of millions of ad nodes, with billions of edges updated daily. We consider the efficiency and scalability in both the SGL and IGL stages:
\textbf{a) The SGL stage.} The embedding algorithm should scale linearly (or better) with graph size, maintain low computational complexity per epoch, and exploit the parallel computing power.
\textbf{b) The IGL stage.} Real-time ad campaigns generate a continuous torrent of new interactions that can quickly reflect changes in user preferences, requiring the synchronized adaptation of embeddings. Rather than retraining from scratch, DHGE must support \textbf{streaming, mini-batch updates} that a) absorb fresh events promptly, b) bound memory growth, and c) guarantee convergence stability under concept drift—all while preserving the quality of existing embeddings.

\textbf{iii) GPU-free real-time IGL.} In practice, low-latency inference is mandatory, yet heavily oversubscribed GPUs often sit in long queues, making on-demand access impossible. Our method is therefore engineered to deliver competitive performance with CPU-only execution, ensuring real-time responsiveness without specialized accelerators.

Although sophisticated methods for DHGE have been proposed after decades of development, they do not fully meet the strict requirements of our recommendation scenarios. In comparison to the previous methods (detailed comparison please refer to Table \ref{tab:comparison}), we address the aforementioned challenges through the following coordinated designs:

\textbf{i) Robust cold-start representation.} We fuse a lightweight graph-transformer-based attribute encoder with an identity embedding layer in the same latent space. We employ a global imputation token for missing features, while the embedding layer uniquely parameterizes nodes via shared ID-type aggregation. Global attention inside the transformer further enables cross-node information transfer even for isolated vertices, while during IGL we enrich behavioral signals via multi-hop BFS sampling—turning data sparsity into an augmentation opportunity.

\textbf{ii) Efficiency \& scalability at scale.} The entire SGL stack is built on SGFormer \cite{wu2023sgformer}, yielding $O(V+E)$ time and $O(V)$ memory—a provable linear bound that readily scales to billion-edge graphs with subgraph sampling via random mini-batch partitioning. The IGL stage utilizes ILLE, whose complexity scales linearly with the incremental data size.

\textbf{iii) stringent GPU-free real-time requirement.} For IGL, our incremental algorithm replaces costly global back-propagation with a matrix-factorization step, avoiding full-graph gradient descent by restricting updates strictly to the incremental data and its local neighborhood. This keeps CPU-only updates within millisecond latency, satisfying the GPU-free requirement in production.

In a nutshell, our contributions can be summarized as follows:
\begin{itemize}
    \item \textbf{Production-Grade System and Workflow:} We design and deploy a practical billion-scale dynamic heterogeneous-graph embedding system and workflow for real-time recommendation. It achieves real-time adaptation to evolving user-item interactions while handling cold-start users/items under strict latency and privacy constraints.
    \item \textbf{A Novel and Scalable Heterogeneous Graph Transformer (HetSGFormer):} A new architecture that models large-scale heterogeneous graphs with linear scalability. HetSGFormer outperforms strong baselines offline and, in live A/B tests, achieved significant lifts in Advertiser Value: 6.11\% vs. TransD, 3.64\% vs. HGT, and 1.43\% vs. PinSAGE. These results confirm HetSGFormer's ability to drive tangible profitability in real-world business scenarios.
    \item \textbf{A Practical GPU-Free Incremental Learning Algorithm (ILLE):} A lightweight, model-disentangled algorithm for rapid, CPU-only incremental updates, crucial for environments with limited GPU resources. Hourly updates with ILLE delivered a \textbf{3.22\%} lift in Advertiser Value and an \textbf{83.2\%} improvement in embedding refresh timeliness over daily retraining.
\end{itemize}

\section{Notation and Problem Definition}

\textbf{Heterogeneous graph.} 
Heterogeneous graphs are distinguished from homogeneous graphs by containing multiple types of objects and relations. In line with HGT \cite{hu2020hgt}, a heterogeneous graph is defined as $G = (\mathcal{V}, \mathcal{E}, \mathcal{T}_{\mathcal{V}}, \mathcal{T}_{\mathcal{E}})$ where:
\begin{itemize}
    \item $\mathcal{V}$ is the node set with $|\mathcal{V}| = V$ nodes.
    \item $\mathcal{T}_{\mathcal{V}} : \mathcal{V} \to \{1,\dots,N_{\text{type}}\}$ is the \textit{node type mapping function}, assigning each node $v \in \mathcal{V}$ to one of the $N_{\text{type}}$ node types.
    \item Each node $v \in \mathcal{V}$ has a feature vector $\mathbf{x}_v \in \mathbb{R}^{D_{\text{in}}}$, forming the node feature matrix $\mathbf{X} \in \mathbb{R}^{V \times {D_{\text{in}}}}$.
    \item $\mathcal{E} \subseteq \mathcal{V} \times \mathcal{V}$ is the edge set with $|\mathcal{E}| = E$ edges.
    \item $\mathcal{T}_{\mathcal{E}} : \mathcal{E} \to \{1,\dots,E_{\text{type}}\}$ is the \textit{edge type mapping function}, assigning each edge to one of the $E_{\text{type}}$ relation types.
    \item Connectivity is represented by the adjacency tensor $\mathbf{A} \in \{0,1\}^{V \times V}$ where $a_{uv} = 1$ iff $(u,v) \in \mathcal{E}$.
\end{itemize}

\noindent We define additional notations that will appear later in the text:
\begin{itemize}
\item For a given type $t$, the set of nodes of type $t$ is $\mathcal{V}_t = \{ v \in \mathcal{V} \mid \mathcal{T}_{\mathcal{V}}(v) = t \}$, and the node count is $V_t = |\mathcal{V}_t|$.
\item Node type vector is $\mathbf{T}_{\text{n}} \in \mathbb{Z}^{V}$, where $\mathbf{T}_{\text{n}}[i] = \mathcal{T}_{\mathcal{V}}(v_i)$ for node $v_i$ at index $i$.
\item Edge type vector is $\mathbf{T}_{\text{e}} \in \mathbb{Z}^{E}$, where $\mathbf{T}_{\text{e}}[i] = \mathcal{T}_{\mathcal{E}}(e_i)$ for edge $e_i$ at index $i$.
\end{itemize}

\noindent\textbf{Graph Embedding.} 
The embedding task learns a mapping function $f: \mathcal{V} \to \mathbb{R}^D$ that projects each node $v \in \mathcal{V}$ to a representation vector $\mathbf{z}_v \in \mathbb{R}^D$, where $D$ is the dimension of embedding and typically $D \ll |\mathcal{V}|$. In the context of heterogeneous graphs, this learned low-dimensional transformation serves as a compact representation that captures both the complex structural connectivity and the diverse type-specific semantics.

\section{Methodology}

\subsection{Overview}
Our framework is built around two tightly coordinated yet fully decoupled modules: a graph transformer that delivers scalable SGL and a model-disentangled dynamic updater that performs lightweight IGL. The remainder of this paper is organized as follows:

\begin{description}
    \item[Sec. \ref{sec:data_preprocessing}] presents a preprocessing method that separates homogeneous from heterogeneous graph fragments, preparing the data for efficient ingestion.
    \item[Sec. \ref{sec:basemodel}] introduces type-specific embeddings and edge-aware attention heads that jointly encode the rich heterogeneity of nodes and relations.
    \item[Sec. \ref{sec:incremodel}] tackles the cold-start of sparse newcomers with a BFS-guided neighbor sampler and an LLE-inspired reconstruction objective, allowing embeddings to evolve on-the-fly.
    \item[Sec. \ref{sec:workflow}] describes the workflow between the GPU-based static trainer and the CPU-based incremental updater, balancing periodic full-graph training with frequent lightweight updates.
    \item[Sec. \ref{sec:complexity}] concludes with a rigorous time-complexity analysis, underscoring the method's readiness for real-world deployment.
\end{description}

\begin{figure*}[!tb]
    \centering
    \includegraphics[width=\textwidth]{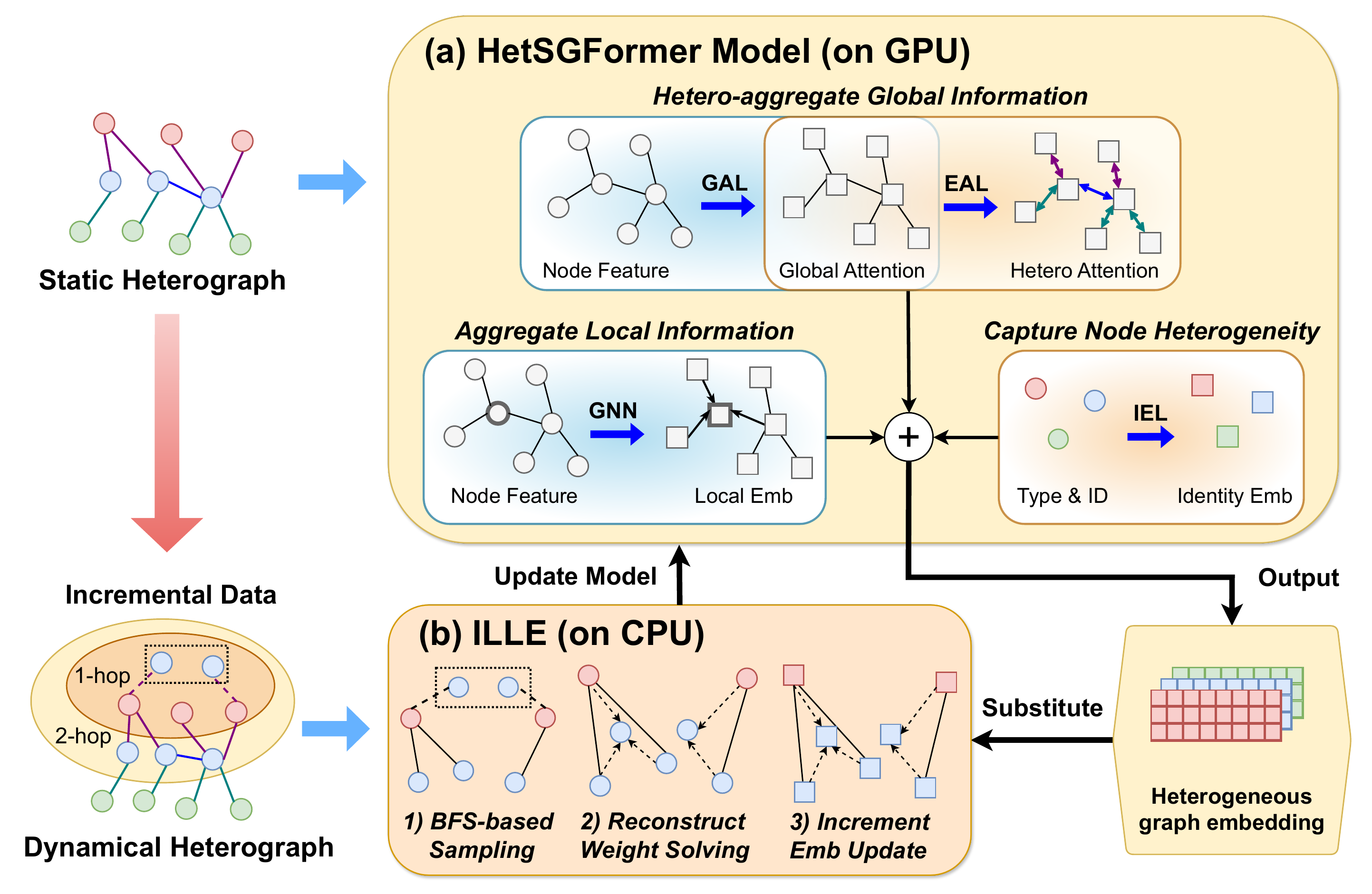}
    \caption{The workflow of the proposed HetSGFormer and ILLE collaboration. The workflow operates as a feedback loop, with frequent CPU updates providing real-time freshness and periodic GPU retraining ensuring the accuracy, timeliness and stability of the model representation. (a) \textbf{HetSGFormer} for the SGL stage. The homogeneous part (highlighted in \textcolor{cyan}{cyan box}) is encoded in the GAL and a GNN module. The GAL captures long-range dependencies across the entire graph, and the GNN accommodates the prior information to generate local embeddings. The heterogeneous part (highlighted in \textcolor{orange}{orange box}) is embedded using the EAL and the IEL module. The EAL aggregates heterogeneous edge information by type-specific edge aggregation, and the IEL transforms heterogeneous node types and identifiers into identity embeddings. (b) \textbf{ILLE} for the IGL stage. For dynamically added nodes and edges, ILLE uses a BFS-based strategy to sample one-hop and two-hop neighbors. The embeddings for these incremental nodes are then reconstructed by linearly aggregating their neighbors' representations, followed by updating the embeddings via a reconstruction loss function.}
    \label{fig:model}
\end{figure*}

\subsection{Graph Preprocessing for Scalability}\label{sec:data_preprocessing}

\noindent \textbf{Subgraph Sampling.} For large-scale heterogeneous graphs, computational constraints necessitate the use of efficient subgraph sampling strategies. We perform neighbor subgraph sampling via random mini-batch partitioning of source nodes \cite{wu2023difformer, wu2023sgformer}, and restrict the maximum neighbor count to a pre-defined constant in each subgraph. This approach not only adapts seamlessly to graphs of varying scales but also allows flexible batch-size adjustment tailored to available hardware resources, with negligible overhead. In subsequent sections, we do not strictly differentiate between the complete graph and a sampled subgraph, as the model treats both inputs consistently.

\noindent \textbf{Input Initialization.} Given a heterogeneous graph $G = (\mathcal{V}, \mathcal{E}, \mathcal{T}_{\mathcal{V}}, \mathcal{T}_{\mathcal{E}})$, we extract the node type vector $\mathbf{T}_{\text{n}}$, raw features $\mathbf{X}$, edge type vector $\mathbf{T}_{\text{e}}$, and the adjacency structure $\mathbf{A}$. We first project the raw features $\mathbf{X}$ into a latent space $\mathbf{X}_0 \in \mathbb{R}^{V\times D}$ via a unified single-layer MLP. To robustly handle missing features in sparse industrial data, we introduce a global learnable imputation token. This design is more parameter-efficient than maintaining type-specific defaults and offers superior generalization for sparse or cold-start nodes.

\subsection{HetSGFormer: \underline{Het}erogeneous \underline{S}ingle-Layer \underline{G}raph Trans\underline{former}}\label{sec:basemodel}

We build HetSGFormer on SGFormer \cite{wu2023sgformer}, which efficiently represents large-scale graphs with linear time complexity and achieves impressive performance on heterogeneous graph mining tasks. While SGFormer lacks the capability to model heterogeneous graphs, our HetSGFormer leverages its efficiency and proposes new modules to capture and aggregate information according to multiple node and edge types, making it a suitable model for the SGL module of large-scale heterogeneous graphs. HetSGFormer comprises two main parts and four key modules. The homogeneous part, inherited from SGFormer, includes the Global Attention Layer and a Graph Neural Network (GNN) module tailored for large-scale graphs; and the heterogeneous part, which we propose, incorporates an Identity Embedding Layer to capture node heterogeneity and an Edge Attention Layer to address edge heterogeneity. In Fig. \ref{fig:model}(a), we show how these modules collaborate with each other.

\noindent \textbf{Identity Embedding Layer (IEL).}
To capture the unique latent semantics and collaborative signals of each entity \cite{he2017neural, he2020lightgcn}, we adopt an intra-type indexing strategy: for each node type $t \in \{1, \dots, N_{\text{type}}\}$, nodes are enumerated individually from $1$ to $V_t$. Under this scheme, identifiers are unique only within their respective types, meaning the same integer index $k$ may refer to distinct entities across different types.

To resolve this ambiguity while optimizing for scalability, we decouple the node representation into two components: a shared identifier embedding and a type-specific embedding. The final representation is obtained by aggregating these two vectors. This design is crucial for handling billion-scale graphs as it significantly reduces memory constraints. Instead of maintaining a distinct embedding for every unique node in the global space which would require a table size of $|\mathcal{V}| \times D$, we utilize a shared identifier table sized at $V_{\text{max}} \times D$, where $V_{\text{max}} = \max_{t} V_t$, alongside a negligible type table of size $N_{\text{type}} \times D$. Since $V_{\text{max}} \ll |\mathcal{V}|$, this parameter-sharing mechanism effectively lowers the memory footprint for large-scale deployment.

\noindent \textbf{Global Attention Layer (GAL).}
In large-scale heterogeneous graphs, nodes often exhibit semantic similarities or feature correlations even when they belong to different types or are distant in the graph topology. To effectively model such cross-node interactions and learn global attentions from node features across the entire graph, we adopt an all-pair attention unit. This unit enables the model to efficiently capture implicit dependencies.

Formally, we first project the initial node features $\mathbf{X}_0$ into query, key, and value spaces using learnable weight matrices $\mathbf{W}^Q, \mathbf{W}^K, \mathbf{W}^V \in \mathbb{R}^{D \times D}$. To ensure numerical stability, we apply Frobenius normalization: $\mathbf{Q} = \frac{\mathbf{X}_0 \mathbf{W}^Q}{\left\|\mathbf{X}_0 \mathbf{W}^Q\right\|_F}$, $\mathbf{K} = \frac{\mathbf{X}_0 \mathbf{W}^K}{\left\|\mathbf{X}_0 \mathbf{W}^K\right\|_F}$, and $\mathbf{V} = \mathbf{X}_0 \mathbf{W}^V$.

To circumvent the prohibitive $O(V^2)$ computational cost of standard Softmax attention \cite{vaswani2017attention}, we adopt a linear attention formulation \cite{wu2023sgformer}. This efficient approach reduces complexity to linear time, ensuring scalability for large graphs. The attention mechanism incorporates a normalization term encoded in a diagonal matrix $\mathbf{D} \in \mathbb{R}^{V \times V}$, defined as:
\begin{equation}\mathbf{D} = \text{diag}\left(\mathbf{1}_V + \frac{1}{V} \mathbf{Q} \left(\mathbf{K}^\top \mathbf{1}_V\right)\right),
\end{equation}
where $\mathbf{1}_V$ denotes a $V$-dimensional vector of ones.

Finally, the GAL aggregates global context while preserving local node information through a weighted residual connection. The final representation is computed as:
\begin{equation}
\label{eq:gal}\text{GAL}(\mathbf{X}) = \beta \mathbf{D}^{-1} \left( \mathbf{V} + \frac{1}{V} \mathbf{Q} \left(\mathbf{K}^\top \mathbf{V}\right) \right) + (1 - \beta) \mathbf{X}_0,
\end{equation}
where $\beta$ controls the trade-off between the global attention update and the original features.

\noindent \textbf{Edge Attention Layer (EAL).}
To effectively encode the structural heterogeneity and semantic dependencies inherent in different edge types, we employ a type-aware attention mechanism. This module aggregates the global context captured by the GAL into refined heterogeneous embeddings.

First, we partition the global representations outputted by GAL according to node types. For each node type $t$, let $\mathbf{G}_t \in \mathbb{R}^{V_t \times D}$ denote the feature matrix. We apply type-specific linear transformations using learnable matrices $\mathbf{W}^Q_t, \mathbf{W}^K_t, \mathbf{W}^V_t \in \mathbb{R}^{D \times D}$ to project these features into query, key, and value spaces: $\mathbf{Q}_t = \mathbf{G}_t\mathbf{W}^Q_t$, $\mathbf{K}_t = \mathbf{G}_t\mathbf{W}^K_t$, and $\mathbf{V}_t = \mathbf{G}_t\mathbf{W}^V_t$. Simultaneously, to capture the semantics of specific relations, we introduce a learnable relationship factor $\mathbf{p}_r$, an attention matrix $\mathbf{R}^{\text{att}}_r$, and a message matrix $\mathbf{R}^{\text{msg}}_r$ for each edge type $r$.

For a relation $r$ connecting a source node type $s$ to a target node type $t$, we compute the edge-specific attention scores as:
\begin{equation}
\hat{\mathbf{A}}_{r} = \text{softmax}\left(\frac{\mathbf{p}_r}{\sqrt{D}} \mathbf{Q}_t \cdot (\mathbf{R}^{\text{att}}_r \mathbf{K}_s^\top)\right).
\end{equation}

Unlike the global all-pair attention in GAL, the EAL operates within each node's local neighborhood. Since the node degree is bounded by a constant limit via subgraph sampling (mentioned in Sec. \ref{sec:data_preprocessing}), we can leverage the high expressivity of Softmax attention with only constant computational complexity \cite{bastos2022softmax}, avoiding the quadratic overhead usually associated with full-graph attention.

Finally, the node representations are updated via residual message passing, aggregating information from all relevant edge types $\mathcal{R}_t$:
\begin{equation}
\mathbf{E}_t (\mathbf{G}_t,\mathcal{R}_t,t) = \beta_t \left( \sum_{r \in \mathcal{R}_t} \hat{\mathbf{A}}_r \left( \mathbf{V}_s \mathbf{R}^{\text{msg}}_r \right) \right) + (1 - \beta_t) \mathbf{G}_t,
\end{equation}
where $\beta_t$ is a type-based learnable parameter balancing the residual connection. To maximize efficiency, these type-specific computations are executed in parallel and subsequently reassembled to match the original node order.

\noindent \textbf{GNN Module.}
Unlike the GAL, which extracts representations from a macro-level global perspective, GNNs focus on local-level structural context. They generate embeddings by aggregating features from local neighborhoods, thereby effectively encoding the geometric priors of the graph structure. This characteristic makes GNNs an ideal component for augmenting Transformer layers, creating a synergy between local connectivity and global relevance \cite{wu2021representing}. In practice, to realize this local encoding, we utilize the Graph Convolutional Network (GCN) \cite{kipf2016semigcn}. The GCN is preferred for its efficiency, which is crucial to our framework and ensures good scalability and rapid convergence even on large-scale graph datasets.

\noindent \textbf{Model training.}
Finally, we aggregate the outputs of the IEL, the EAL and the GNN. For the choice of aggregation, we employ a weighted sum with hyperparameter weights instead of linear projection, as empirical results demonstrate that parameter efficiency accelerates convergence in large-scale graphs, addressing the timeliness requirements of recommendation systems.

We employ an unsupervised loss function for learning by sampling positive and negative node pairs \cite{hamilton2017inductive, ying2018pinsage}. The objective is to maximize the similarity between node embeddings for observed edges (positive pairs) while minimizing similarity for non-edges (negative pairs). Specifically, the loss function is defined as:
\begin{equation}
\mathcal{L} = -\sum_{(i,j) \in \mathcal{E}} \log \sigma(\mathbf{z}_i^T \mathbf{z}_j) - \sum_{(i',j') \in \mathcal{E}_{\text{neg}}} \log \sigma(-\mathbf{z}_{i'}^T \mathbf{z}_{j'}),
\end{equation}
where $\mathcal{E}$ represents the set of existing edges in the graph, $\mathcal{E}_{\text{neg}}$ represents the set of non-existing edges in the graph, $\mathbf{z}_i$ is the embedding vector of node $i$, and $\sigma(x)$ is the sigmoid function. The size of $\mathcal{E}_{\text{neg}}$ is set to match that of $\mathcal{E}$, which empirically yielded the best performance.

To construct $\mathcal{E}_{\text{neg}}$ effectively, we adopt the Dynamic Negative Sampling strategy \cite{zhang2013optimizing}. For each positive pair $(i,j)$, we uniformly sample a pool of candidate negatives from the non-interacted set, evaluate them using the current model, and select the candidate with the highest similarity score $j'$. This approach maintains a 1:1 ratio between positive and negative pairs while ensuring the model focuses on discriminative learning against difficult instances.

\subsection{ILLE: Model-Disentangled \underline{I}ncremental \underline{L}ocally \underline{L
}inear \underline{E}mbedding}\label{sec:incremodel}

In real-world recommendation scenarios, user-item interactions emerge continuously as high-frequency, low-volume data streams. Processing these frequent yet small increments demands a delicate balance between real-time responsiveness and prediction accuracy. To address this, we design the IGL module to be fully decoupled from the SGL module. This architectural separation not only facilitates efficient handling of streaming data but also ensures forward compatibility. The framework remains effective and pluggable even if the underlying SGL backbone is upgraded or replaced with a newer model in future iterations.

We propose ILLE \cite{roweis2000nonlinear, chen2011locally}, an approach for efficiently updating low-dimensional representations of dynamical data. As shown in Fig. \ref{fig:model} (b)), the ILLE algorithm involves three steps: selecting neighborhoods to construct a local graph, computing reconstruction weights from neighbors, and embedding the data points into a lower-dimensional space \cite{kouropteva2005incremental}. Instead of relying solely on LLE, we enhance the framework by replacing costly global back-propagation with a localized matrix-factorization step. This approach restricts updates strictly to the new data and its neighborhood, thereby avoiding the redundancy of full-graph retraining, enabling efficient adaptation to continuous incremental data streams, and providing a scalable solution for dynamic environments. As a model-disentangled method \cite{higgins2017beta}, ILLE facilitates updating specific model components without requiring adjustments to all parameters.

\noindent \textbf{Neighbor Selection.}
To accommodate incremental nodes connected to the existing heterogeneous graph (Fig. \ref{fig:heterograph}, dashed box), we substitute the standard k-nearest neighbors (kNN) selection of LLE with a BFS-based sampling strategy. Specifically, we sample both first-order (one-hop) and second-order (two-hop) neighbors. To maintain a fixed neighborhood size $k$, we employ sampling with replacement if the initial neighbor count is insufficient. This multi-hop design is strategic: first-order neighbors allow new nodes to inherit direct attribute information from connected entities, while second-order neighbors facilitate the aggregation of collaborative signals from same-type nodes, thereby enriching the representation context.

\noindent \textbf{Reconstruction Weights Calculation.}
For each incremental data point, we compute the reconstruction weights based on its local neighborhood by solving the following constrained optimization problem \cite{chen2011locally}:
\begin{equation}
\min \sum_{i=1}^{n} \left\|x_i - \sum_{j=1}^{k} w_{ij} x_{j}\right\|_2^2,  \quad \text{s.t.} \quad \sum_{j=1}^{k} w_{ij} = 1.
\end{equation}
Here, $n$ denotes the number of incremental nodes, $x_i$ represents the feature vector of the $i$-th node, and $x_{j}$ denotes its $j$-th neighbor. Upon obtaining the optimal weights, we perform a smooth update of the node representation via a residual connection: $x_i' = \alpha \sum_{j=1}^{k} w_{ij} x_{j} + (1-\alpha) x_i$, where $\alpha$ is a balancing hyperparameter.

The model-disentanglement capability of ILLE is concretely exemplified in the selection of high-dimensional representations for data points. For example, with HetSGFormer, we only need to update the IEL module for new user or item nodes. This is because the GAL module generates global attention, and the EAL and GNN modules, which aggregate local neighborhoods, have already been well-trained within the base model. Without the need to load all model parameters, this approach allows targeted updates only to modules critical to structural embedding, ensuring adaptability across diverse model architectures while reducing computational demands and improving operational efficiency.

\noindent \textbf{Embedding Generation and training.}
To obtain the final low-dimensional representations, we map the data into a latent space while preserving the local geometric structures captured by the weights. This is achieved by minimizing the embedding reconstruction error \cite{hou2009local, chang2006robust}:
\begin{equation}
\mathcal{L} = \sum_{i=1}^{n} \left\| y_i - \sum_{j=1}^{k} w_{ij} y_j \right\|_2^2,
\end{equation}
where $y_i$ is the embedding corresponding to the high-dimensional node $x_i$.

\subsubsection{Parameter Sensitivity}
The choice of the neighborhood size $k$ introduces a critical trade-off between efficiency and performance. A value of $k$ that is too small may result in insufficient neighborhood sampling, limiting the utilization of local information and hindering the cold-start performance for new nodes. Conversely, an excessively large $k$ not only significantly increases computational overhead but may also introduce data redundancy given the inherent sparsity of recommendation graphs, potentially hampering model convergence. By selecting an appropriate $k$ value, ILLE remains well-suited for efficiently handling dynamic data streams. To empirically determine this optimal balance, we will conduct a detailed sensitivity analysis of $k$ in \ref{sec:offline_experiment}.

To address potential scenarios involving dense graphs or high-degree nodes where the candidate neighbor count exceeds the optimal $k$, our approach can avoid the computational penalty of naively increasing $k$. We maintain a fixed, manageable $k$ by randomly sampling from the neighborhood while explicitly preserving the priority of one-hop and two-hop connections. To ensure comprehensive coverage of the local structure, the algorithm relies on multiple iterative updates to traverse distinct subsets of neighbors over time. By combining an appropriate $k$ value with this sampling strategy, ILLE remains robust and efficient even when handling dense structural patterns in dynamic data streams.

\subsubsection{Further Discussion on ILLE's Incremental Update}\label{sec:incre_update}

To understand ILLE's ability to process only incremental data, we first briefly recall some key conclusions of the standard LLE model.

Given $N$ nodes, LLE computes the embedding $\mathbf{Y} \in \mathbb{R}^{N \times D}$ by solving the eigenvalue problem $\mathbf{Y}^{\top}\mathbf{MY} = \mathbf{\Lambda}$, where $\mathbf{M}=(\mathbf{I-W})^\top (\mathbf{I-W})$ is the alignment matrix and $\mathbf{\Lambda}$ contains the $D$ smallest non-zero eigenvalues. This process requires a computationally expensive global eigendecomposition.

In dynamic settings where $N_t$ new nodes arrive, recomputing the entire spectrum is prohibitive. Instead, ILLE adopts an incremental optimization strategy. Let $\mathbf{Y}' \in \mathbb{R}^{(N+N_t) \times D}$ and $\mathbf{M}'$ denote the augmented embedding and alignment matrices. Relying on the spectral stability assumption that the smallest eigenvalues remain relatively invariant during minor updates ($\mathbf{\Lambda}' \approx \mathbf{\Lambda}$), we transform the intractable eigen-problem into a tractable constrained minimization problem:
\begin{equation}
\min_{\mathbf{Y}'} \mathcal{J} = \left\| \mathbf{Y}'^\top \mathbf{M}' \mathbf{Y}' - \mathbf{\Lambda} \right\|_F^2, \quad \text{s.t.} \quad \mathbf{Y}'^\top \mathbf{Y}' = (N+N_t)\mathbf{I}.
\end{equation}
Instead of solving for eigenvectors directly, we treat this as an optimization of the Frobenius norm on a $D \times D$ matrix. This incremental formulation can be solved efficiently using an interior-point method, therefore avoiding the global decomposition of the large $(N + N_t) \times (N + N_t)$ matrix. By initializing $\mathbf{Y}'$ with the existing embeddings and applying gradient descent, we update the representations of the incremental nodes and their immediate neighbors. This reduces the complexity from performing eigendecomposition on a full-graph matrix to optimizing a low-dimensional objective, significantly lowering the computational overhead for real-time streams.

\subsection{Hybrid CPU-GPU Collaboration and Workflow}\label{sec:workflow}

To balance the efficiency of dynamic updates with the ability to learn from large-scale data, we design a collaborative workflow between HetSGFormer and ILLE, operating across GPU and CPU environments. The detailed interaction is illustrated in Fig. \ref{fig:model}.

\textbf{Static Phase on GPU.} The HetSGFormer module is executed on the GPU to process existing large-scale static heterogeneous graph data. As shown in Fig. \ref{fig:model}(a), the homogeneous part of the graph is embedded using the GAL and a GNN module, while the heterogeneous part utilizes the IEL and EAL modules to transform node types and aggregate edge information. This phase involves mini-batch subgraph sampling and full-graph gradient descent to train the \textbf{static model}, outputting comprehensive graph embeddings for all nodes in the static graph.

\textbf{Incremental Phase on CPU.} As new users, items, and interactions are progressively incorporated, the system shifts focus exclusively to the neighborhoods of these incremental nodes. This process is handled by the ILLE algorithm solely on the CPU. As depicted in Fig. \ref{fig:model}(b), for dynamically added nodes and edges, we employ a BFS-based strategy to sample one-hop and two-hop neighbors. ILLE generates embeddings for these incremental nodes via linear reconstruction, aggregating the embeddings of sampled neighbors. The hybrid collaboration is achieved through a feedback loop. The embeddings reconstructed by the CPU are fed back to the static model to update its parameters, resulting in an \textbf{incremental model}.

\textbf{Iterative Update Workflow.} In real-world applications, since dynamic graphs update more frequently than static graphs, we treat this incremental model as the reference for subsequent ILLE iterations. Our framework prioritizes retaining historical data to build long-term interest profiles. Instead of performing explicit real-time deletions, which would erase user history, we rely on periodic static graph reconstruction to handle outdated behaviors or implicitly apply time-decay strategies. This circulation on ILLE continues until the static graph is updated, at which point HetSGFormer trains a new static model on the GPU. Under this design, resource-intensive GPU training is required only during periodic static graph updates, whereas frequent incremental updates are efficiently handled on the CPU.

\subsection{Complexity Analysis}\label{sec:complexity}

The overall computational complexity of HetSGFormer is $O(V+E)$. This arises because the GAL utilizes linear attention, while the IEL requires $O(V)$ and the GCN module requires $O(E)$ operations. Although the theoretical time complexity of the EAL is non-linear, leveraging graph sparsity and employing a degree-limited neighbor sampling strategy effectively reduces it to $O(V)$.

For ILLE, the overall computational complexity depends on the number of updated data points $N_{\text{upd}}$ and scales as $O(N_{\text{upd}} \times k^3)$. This complexity stems from three main operations: neighbor search requiring $O(N_{\text{upd}})$ operations, reconstruction weight calculation costing $O(N_{\text{upd}} \times k^3)$ operations, and updating the affected nodes involving $O(k^3)$ operations.

\section{Experiments}

\subsection{Overview}

To comprehensively validate the practicality, generalization, and commercial value of our HetSGFormer+ILLE framework, we conducted multi-dimensional experiments spanning private industrial data, public benchmarks, and real-world commercial deployment. Offline evaluations on a massive private dataset (53M users, 3M ads, 158M interactions) verified its robustness in handling cold-start and scalability challenges, outperforming baselines like TransD, PinSAGE, and HGT on Recall@K and HitRate@K. Complementary tests on four public datasets (Ali-Display, Epinions, Amazon-CD, Yelp) confirmed its strong generalization, achieving top performance in 13 out of 16 (dataset, metric) combinations. Most critically, online A/B tests on a billion-scale commercial advertising platform demonstrated tangible business impact: HetSGFormer lifted Advertiser Value by up to 6.11\%, while ILLE added a further 3.22\% lift and improved embedding refresh timeliness by 83.2\%. These results collectively validate the framework’s technical reliability and successful production deployment, translating model advancements into measurable revenue growth.

\noindent \textbf{Privacy Claim.} To validate the proposed approach in a real-world setting, we utilized a private industrial dataset derived from the advertising billing logs of an international company for offline evaluation. Furthermore, we conducted rigorous online A/B testing on the same company's platform. However, adherence to strict commercial confidentiality constraints is mandatory. As a result, specific sensitive details cannot be disclosed. Specifically, the exact model hyperparameter settings, the raw data distributions beyond aggregate statistics, and the absolute monetary figures for Advertiser Value in online experiments remain confidential. Consequently, only relative performance lifts and high-level descriptions are reported.

\subsection{Offline Experiments On Private Datasets}\label{sec:offline_experiment}

\subsubsection{Settings}\label{sec:offline_experiment_setting}

\textbf{Experimental Setup.} We conduct offline evaluations using advertising billing data from the private dataset to evaluate the model's ranking capability and retrieval accuracy. To this end, the data is partitioned into training and test sets based on the chronological sequence, with user click-through behavior indicating interest. Performance is assessed by comparing recommended items against ground-truth user interaction sequences.

In dynamic scenarios, our process involves updating embeddings of modified and newly added nodes. These updated embeddings are then used to generate personalized recommendations. To evaluate dynamic adaptation, we select the optimal baseline model as the reference for incremental experiments. The performance of the ILLE-updated model is benchmarked against this reference, quantifying performance differences and providing insights into the model's adaptability in evolving environments. Hyperparameter settings are omitted due to commercial secrets.

\noindent \textbf{ILLE Paramenter Settings.} For the ILLE module, the neighborhood size $k$ serves as a critical hyperparameter governing the scope of local information aggregation. To identify the optimal configuration that balances structural context with noise avoidance, we conducted a sensitivity analysis by varying $k$ within the set $\{6, 8, 10\}$. Based on the performance validation (detailed in Sec. \ref{sec:offline_results}), we adopt $k=8$ as the default setting for all subsequent main experiments.

\noindent \textbf{Baselines.}
We compared our HetSGFormer+ILLE with the knowledge graph embedding model TransD \cite{ji2015transd}, the GNN-based model PinSAGE \cite{ying2018pinsage}, and the graph transformer model HGT \cite{hu2020hgt}, all of which are or have been deployed in mainstream commercial applications. We will introduce them in Sec. \ref{sec:related}.

\noindent \textbf{Ablations.}
To systematically validate the contributions of HetSGFormer's multi-view architecture, we designed a comprehensive set of ablation variants. It is crucial to note that the vanilla SGFormer architecture integrates the GAL module with a GNN to combine global and local interactions. However, as a homogeneous encoder, it inherently lacks an Identity Embedding Layer (IEL). In real-world recommendation scenarios, capturing unique user and item identities is fundamental. Without IEL, the vanilla SGFormer is practically ill-suited for such tasks.

To address this and enable a meaningful comparison, we align our ablation baselines to the SGFormer architecture adapted for heterogeneous data. We compare the full \textbf{HetSGFormer} and its incremental variant \textbf{HetSGFormer+ILLE} against the following versions:

\begin{itemize}
\item \textbf{HetSGFormer w/o EAL:} This variant removes the EAL. Excluding the heterogeneous edge view underscores the importance of aggregating edge information by specific edge types to capture diverse interaction patterns between endpoints.
\item \textbf{HetSGFormer w/o IEL:} This variant removes the IEL. By excluding the heterogeneous identity view, it highlights the critical role of ID embeddings in recommendation. Since node features alone are often insufficient for precisely identifying distinct users, ID embeddings are indispensable for capturing powerful collaborative signals and unique preferences.
\item \textbf{HetSGFormer w/o GNN:} This variant removes the GNN module. Excluding the local homogeneous view confirms the necessity of accommodating structural prior information and generating robust local embeddings.
\item \textbf{HetSGFormer only IEL:} This variant retains strictly the IEL. It demonstrates that relying solely on ID embeddings is inadequate, as it fails to leverage the rich heterogeneous contexts within the graph—such as node features, global correlations, and edge structures—that are essential for comprehensive representation learning.
\end{itemize}

\noindent \textbf{Datasets.}
User click-through behavior and relevant features are extracted from real billing records. For base model selection, three consecutive days of data are partitioned for training. This dataset encompasses 53 million unique users, 3 million unique ads, and 158 million clicks, a scale that significantly stresses both model training and incremental updating capabilities. The data from the subsequent day is segmented so that the initial hours are allocated to incremental model training in dynamic scenarios, which processes tens of thousands of modified or added items per hour. The latter hours are reserved for performance evaluation.

\noindent \textbf{Computational Resource.}
The base model was trained on a computing cluster featuring a 16-core CPU, 128 GB of RAM, and dual Tesla V100 GPUs. The incremental model used an identical CPU/RAM configuration (16 cores, 128GB) but deliberately excluded GPU acceleration. This resource-constrained design simulates real-world environments where GPU scarcity weakens computational capacity during model updates. 

\noindent \textbf{Metrics.}
\textbf{HitRate@K} and \textbf{Recall@K} are used to evaluate the offline model's performance. HitRate@K measures whether relevant items appear among the top $K$ recommendations, reflecting the model's precision in capturing user interest within the limited top-ranked list. Recall@K measures the proportion of relevant items found in the top K recommendations, reflecting the model's ability to retrieve relevant content. Both metrics are is defined as:
\begin{align}
  \mathrm{HitRate}@K &= 
  \frac{1}{\lvert \mathcal{U}_{\mathrm{test}} \rvert}
  \sum_{u \in \mathcal{U}_{\mathrm{test}}} \mathbb{I}\big[i_u \in \mathcal{R}^K_u\big],\\
  \mathrm{Recall}@K &= 
  \frac{1}{\lvert \mathcal{U}_{\mathrm{test}} \rvert} 
  \sum_{u \in \mathcal{U}_{\mathrm{test}}} 
  \frac{\lvert \mathcal{I}_{u} \cap \mathcal{R}^K_u \rvert}{\lvert \mathcal{I}_{u} \rvert}, 
\end{align}
where $\mathcal{U}_{\mathrm{test}}$ is the set of test users, $i_u$ is the relevant items for user $u$, $\mathcal{R}^K_u$ is the top-$K$ recommendation list for user $u$, and $\mathcal{I}_{u}$ is the ground-truth item set for user $u$.

In our research, considering the scale of the heterogeneous graph, we set K as 256, 512, and 1024.

\subsubsection{Recommendation Lists Generation in Serving}\label{sec:ANN}
In the deployment phase, the model outputs graph embeddings for all nodes, integrating representations from both the static heterogeneous graph and incremental data streams. These learned user and item embeddings are used to quantify user interest via cosine similarity. During user interactions with applications and services provided by the server, recommendation lists are efficiently generated via cosine similarity-based Approximate Nearest Neighbor (ANN) search \cite{arya1998optimal}. The application of ANN constitutes a standard practice in billion-scale recommendation systems \cite{johnson2017fbann}. This same approach is also utilized in Sec. \ref{sec:online_experiment}.

\subsubsection{Results}\label{sec:offline_results}

\begin{figure}[!tb]
    \centering
    \includegraphics[width=\columnwidth]{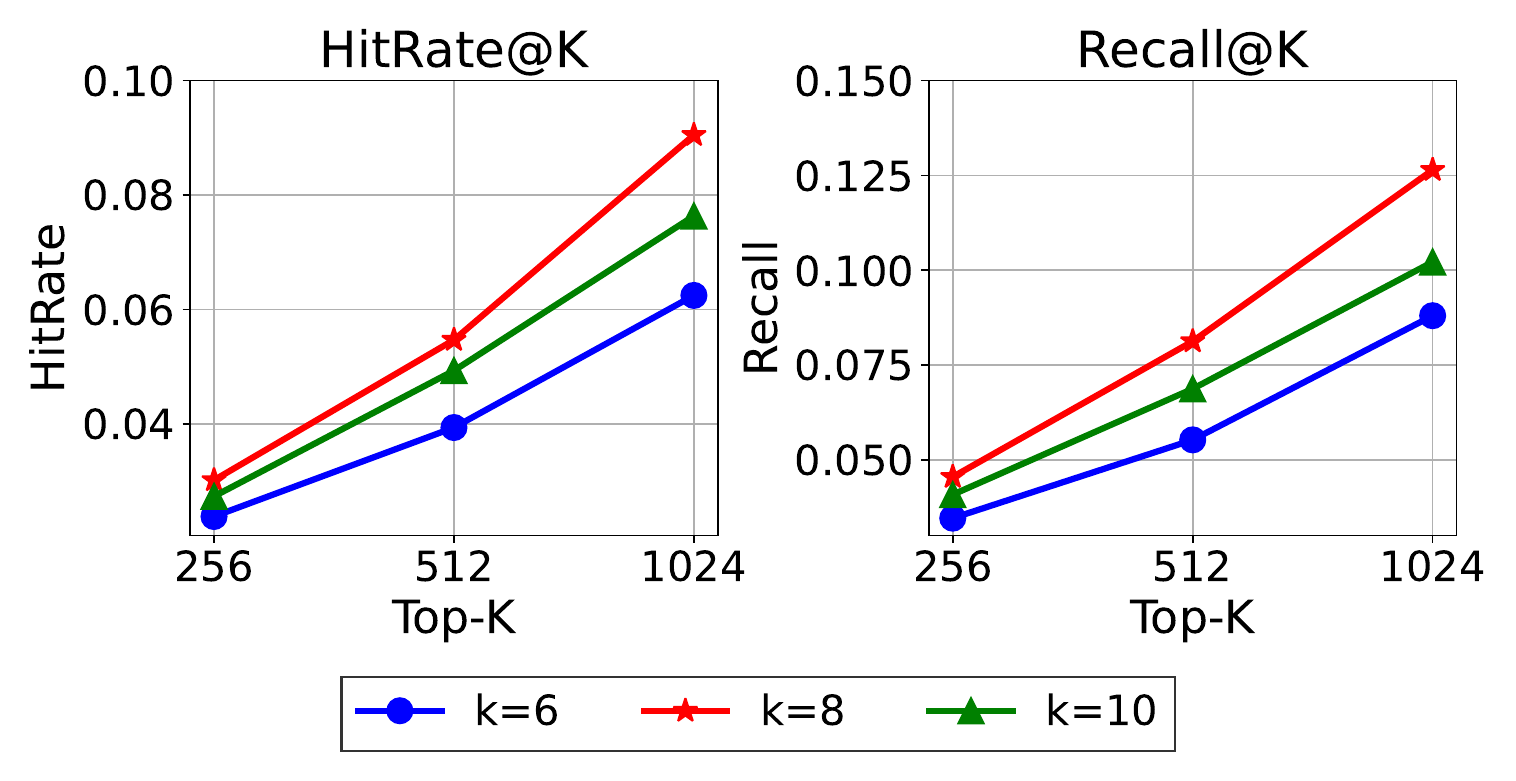}
    \caption{Evaluation results showing the impact of the neighborhood size parameter $k$ on the performance of ILLE. Evaluated using HitRate@K and Recall@K metrics across different retrieval sizes.}
    \label{fig:parameter_experiment}
\end{figure}

\textbf{Parameter sensitivity.}
The impact of the neighborhood size $k$ on model performance is illustrated in Fig. \ref{fig:parameter_experiment}. As observed, increasing $k$ from 6 to 8 yields a noticeable performance gain across all metrics. This improvement demonstrates that an adequately sized neighborhood is essential for capturing sufficient structural context to effectively reconstruct embeddings for new nodes, particularly to alleviate the cold-start problem. However, performance degrades when $k$ is further increased to 10. This decline aligns with our analysis that, given the inherent sparsity of recommendation graphs, an excessively large $k$ forces the model to incorporate irrelevant or distant neighbors. This introduces noise and data redundancy, which hampers model convergence. Therefore, $k=8$ is identified as the optimal point, balancing information sufficiency and noise robustness.

\begin{figure}[!tb]
    \centering
    \includegraphics[width=\columnwidth]{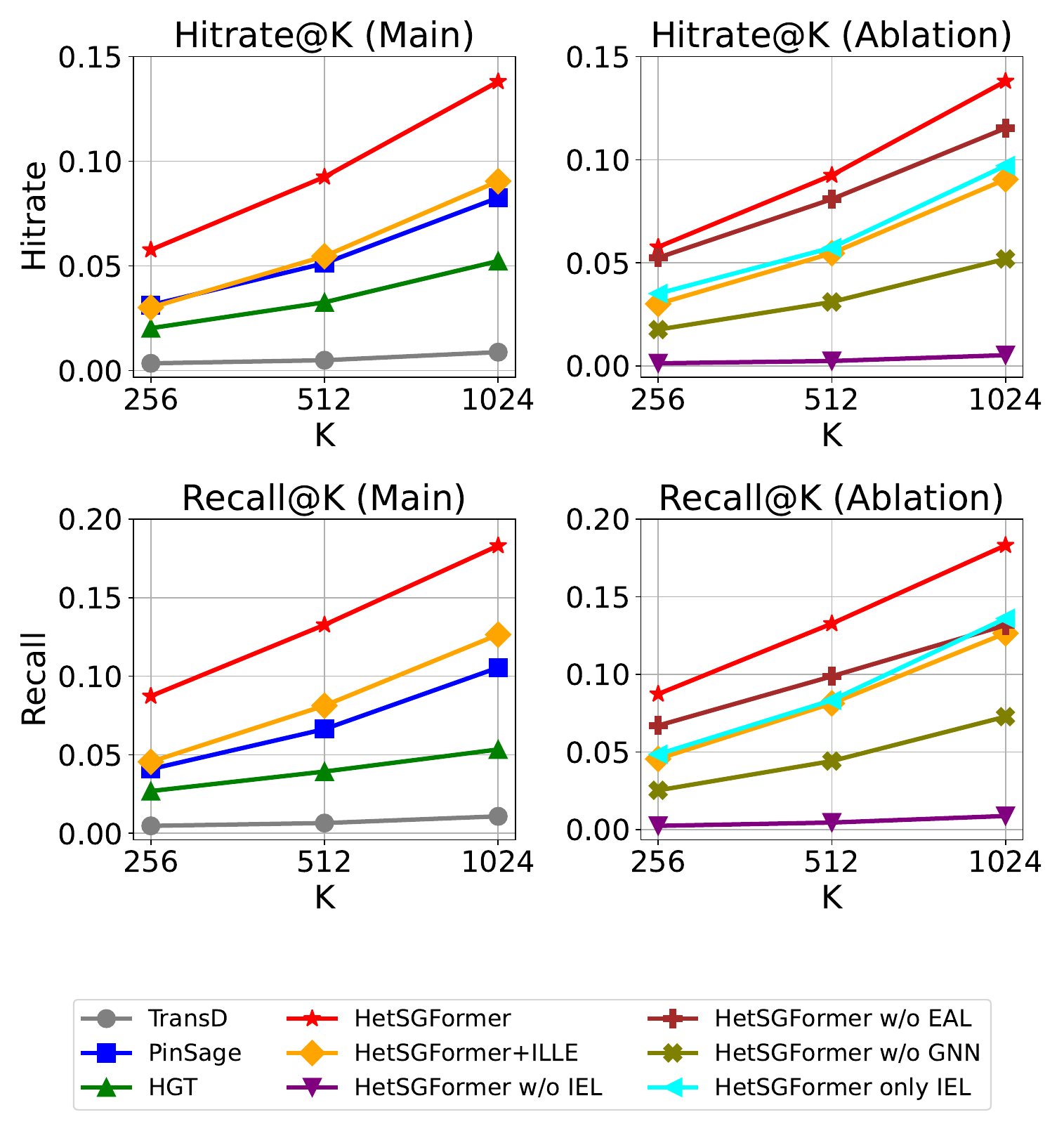}
    \caption{Evaluation results comparison of three baseline models, our HetSGFormer) and ILLE-enhanced HetSGFormer on private datasets, with ablation studies on HetSGFormer components. Evaluated using HitRate@K and Recall@K.}
    \label{fig:offline_experiment}
\end{figure}

\noindent \textbf{Model comparison.}
As shown in Fig. \ref{fig:offline_experiment}, the results demonstrate that HetSGFormer significantly outperforms all baseline models on both metrics and consistently achieves superior performance compared to TransD, PinSAGE, and HGT across all evaluated $K$ values. This significant advantage highlights HetSGFormer's ability to model large-scale heterogeneous relationships within the graph structure, validating its effectiveness for recommendation tasks.

While the ILLE-enhanced variant exhibits a modest performance reduction compared to the original model, it remains highly effective, surpassing strong baselines such as PinSAGE. Crucially, ILLE introduces dynamic updating capabilities that allow continuous model adaptation to evolving graph data. Consequently, the minor performance trade-off is acceptable given the critical need for fresh recommendations in real-world systems.

\noindent \textbf{Ablation study.}
\label{sec:online_ablation}
Presented in Fig. \ref{fig:offline_experiment}, ablation studies systematically validate the contributions of HetSGFormer's multi-view architecture. The significant underperformance of the version that only contains the IEL variant underscores the inadequacy of pure ID embeddings, which fail to leverage rich node features and heterogeneous contexts. These results also highlight the necessity of the GAL’s global perspective on all-pair relativeness and the EAL’s modeling of edge heterogeneity. Furthermore, the observed performance degradation without GNN confirms that it provides a vital local neighborhood perspective, effectively augmenting the Transformer layers with structural information. Collectively, these findings demonstrate that integrating global, edge-specific, and local views is indispensable for the model’s ability to process heterogeneous data and maintain optimal recommendation quality.

\subsection{Online Experiments}\label{sec:online_experiment}
\subsubsection{Settings}

\textbf{Experimental Setup.} Following the offline evaluation, we further verified the model's capability to drive business growth through an online test deployed on the same company's advertising platform. Specifically, we compared three distinct retrieval paths—vector, inverted index, and our graph-based approach—to assess their respective performance in capturing user interests and optimizing revenue-related metrics in real-time recommendation scenarios. Computational resources settings keep the same as offline experiments in Sec. \ref{sec:offline_experiment_setting}. Recommendation lists are still generated by ANN, as described in Sec. \ref{sec:ANN}. 

The primary objective is to evaluate the real-world commercial impact of our dynamic heterogeneous graph embedding method against established standards. We benchmark HetSGFormer against previously established baselines in Sec. \ref{sec:offline_experiment_setting}, including TransD, PinSAGE, and HGT. Furthermore, to substantiate the commercial effectiveness of our ILLE approach in the recommendation system, we deployed a controlled online A/B experiment. The control group used HetSGFormer for graph retrieval, undergoing a full daily retraining to incorporate the latest user behavior and new items. The treatment group is based on HetSGFormer with daily full retraining, but additionally integrates the ILLE algorithm hourly to incrementally update embeddings for modified or newly added nodes, as mentioned in Sec. \ref{sec:workflow}. 

\noindent \textbf{Datasets.}
While the daily volume remained consistent with the massive statistics described in Sec. \ref{sec:offline_experiment_setting}, as the experiment duration increased, the cumulative data volume grew proportionally to the runtime. Consequently, all online tests are operated on a massive heterogeneous graph spanning billions of edges. Due to commercial confidentiality constraints, further sensitive implementation details cannot be disclosed.

\noindent \textbf{Metrics.}
The primary evaluation metric is \textbf{Advertiser Value}, which directly quantifies advertiser Return on Investment. Maximizing this metric fosters sustained platform investment and ecosystem growth by aligning platform incentives with advertiser success. This metric is calculated as:
\begin{equation}
  \text{Advertiser Value} =  \text{\# Impressions} \times \text{CTR} \times \text{CVR} \times \text{Avg Bid},
\end{equation}
where Impressions refer to the number of times a piece of content, such as an advertisement or product, is displayed to a user, CTR (Click-Through Rate) measures the proportion of users who click on the displayed content after viewing it, CVR (Conversion Rate) indicates the percentage of users who complete a desired action, such as making a purchase, after clicking, Avg Bid refers to the average bid price for the advertisement.

To explicitly validate the proposed ILLE's model update efficiency, complementary efficiency metrics are concurrently evaluated, with \textbf{embedding refresh latency} defined as the time interval between consecutive model updates. This metric measures model updating speed, reflecting the timeliness of the recommendation system. It indicates how quickly the system can adapt to new data, which is crucial for maintaining the relevance and accuracy of the recommendations over time.

\subsubsection{Results}

\textbf{Results for HetSGFormer.} We evaluated HetSGFormer against three baselines to ensure a robust comparison. The model demonstrated consistent superiority, achieving lifts in Advertiser Value of \textbf{6.11\%}, \textbf{3.64\%}, and \textbf{1.43\%} compared to TransD, HGT, and PinSAGE, respectively. These metrics specifically reflect the model's profitability in real-world business scenarios, validating its commercial effectiveness.  Due to commercial confidentiality, the specific results cannot be displayed.

\noindent \textbf{A/B test for ILLE.} 
The treatment group exhibited a \textbf{3.22\%} lift in Advertiser Value against the control group. Since the model architecture remains consistent across the two groups, this performance gain is directly attributed to ILLE's ability to optimize parameters using incremental data, rather than to any degradation of the baseline. Additionally, an \textbf{83.2\%} increase in embedding refresh latency was observed compared to the control group. This result demonstrates the efficacy and swift adaptation of our ILLE-enhanced dynamic heterogeneous graph embedding method in leveraging newly added data and delivering more relevant and timely recommendations. For privacy reasons, we cannot show the specific results.

\noindent \textbf{Efficiency analysis.}
We further evaluate the computational efficiency and system resource consumption of the proposed framework. Regarding the SGL stage, HetSGFormer requires 5.75h to complete, which is significantly more efficient than complex GNN baselines like PinSAGE (7.17h) and HGT (9.46h), though naturally slower than the lightweight TransD (1.33h). However, the critical efficiency gain stems from the IGL stage. ILLE reduces the computational cost to merely 0.21h per update round, avoiding the extensive time required for full model retraining. Furthermore, during operation, ILLE demonstrates low resource consumption, with CPU and memory usage ranging between 25\% and 40\%. These results show that the inference latency and memory usage of our model are comparable to those of several simpler baseline models, and confirm that our approach effectively balances rapid model iteration with low resource consumption, validating its high scalability for real-world deployment.

\begin{table}[!tb]
\centering
\caption{Statistics of the public datasets. Sparsity is calculated as $1 - \frac{\#\text{Interactions}}{\#\text{Users} \times \#\text{Items}}$.}
\label{tab:dataset_stats}
\resizebox{\columnwidth}{!}{%
\begin{tabular}{lcccccc}
\toprule
\textbf{Dataset} & \# Users & \# Items & \# Interactions & \textbf{Sparsity} \\
\midrule
Ali-Display & 17,730 & 10,036 & 173,111 & 99.9027\% \\
Epinions & 15,210 & 233,929 & 630,391 & 99.9823\% \\
Amazon-CD & 51,266 & 46,463 & 731,734 & 99.9692\% \\
Yelp & 167,037 & 79,471 & 1,970,721 & 99.9851\% \\
\bottomrule
\end{tabular}%
}
\end{table}

\subsection{On Public Benchmarks}

\begin{table*}[!ht]
  \centering
  \scriptsize
  \caption{\textbf{Public Benchmarks evaluation results on four recommendation datasets. We categorize baselines into Traditional GCNs, Graph Transformers, and Heterogeneous Methods to highlight the comparative advantages of our Proposed HetSGFormer.}}
  \label{tab:metrics_comparison_updated}
  \resizebox{\textwidth}{!}{%
  \begin{tabular}{llrrrrrrrrrrrrr}
    \toprule
    \multirow{2}{*}{Dataset} & \multirow{2}{*}{Metric} & \multicolumn{2}{c}{Traditional GCNs} & \multicolumn{2}{c}{Graph Transformers} & \multicolumn{2}{c}{Heterograph Methods} & \multicolumn{2}{c}{Our Proposed} \\
    \cmidrule(lr){3-4} \cmidrule(lr){5-6} \cmidrule(lr){7-8} \cmidrule(lr){9-10}
    & & LightGCN & PinSAGE & DiFFormer & GFormer & HGT & HERec & HetSGFormer & HetSGFormer + ILLE \\
    \midrule
    \multirow{4}{*}{Ali-Display} 
    & HitRate@10 $\uparrow$   & 0.6982 & 0.7356 & 0.5783 & 0.7050 & 0.6939 & 0.7053  & \textbf{0.7492} & 0.6737  \\
    & NDCG@10 $\uparrow$ & 0.5108 & 0.4907 & 0.3964 & 0.5389 & 0.5430 & 0.5165 & \textbf{0.5673} & 0.5155 \\
    & HitRate@20 $\uparrow$    & 0.7421 & 0.7814 & 0.6296 & 0.7499 & 0.7425 & 0.7541 & \textbf{0.8013} & 0.7418 \\
    & NDCG@20 $\uparrow$  & 0.4983 & 0.4832 & 0.4897 & 0.5781 & 0.5112 & 0.5408 & \textbf{0.6016} & 0.5614 \\
    \midrule
    \multirow{4}{*}{Epinions} 
    & HitRate@10 $\uparrow$    & 0.7205 & 0.8056 & 0.6892 & 0.8045 & 0.7701 & 0.7459 & \textbf{0.8072} & 0.7329 \\
    & NDCG@10 $\uparrow$  & 0.6143 & 0.5381 & 0.4002 & 0.5907 & 0.6539 & 0.6214 & \textbf{0.6682} & 0.6203 \\
    & HitRate@20 $\uparrow$    & 0.7658 & 0.8265 & 0.6313 & \textbf{0.8522} & 0.8116 & 0.7541 & 0.8493 & 0.7736 \\
    & NDCG@20 $\uparrow$  & 0.5824 & 0.5174 & 0.5273 & 0.5788 & 0.6194 & 0.6037 & \textbf{0.6256} & 0.5991 \\
    \midrule
    \multirow{4}{*}{Amazon-CD} 
    & HitRate@10 $\uparrow$   & 0.7168 & 0.8221 & 0.6044 & 0.7321 & 0.7955 & 0.7833 & \textbf{0.8156}  & 0.7375  \\
    & NDCG@10 $\uparrow$ & 0.5522 & 0.6324 & 0.5419 & 0.5597 & 0.6365 & 0.6210 & \textbf{0.6387} & 0.5728 \\
    & HitRate@20 $\uparrow$   & 0.7876 & 0.8917 & 0.6499 & 0.7965 & 0.8604 & 0.8381 & \textbf{0.8923}  & 0.8031  \\
    & NDCG@20 $\uparrow$ & 0.5487 & 0.6251 & 0.5396 & 0.5702 & 0.6304 & 0.6357 & \textbf{0.6489} & 0.5960 \\
    \midrule
    \multirow{4}{*}{Yelp} 
    & HitRate@10 $\uparrow$    & 0.8043 & 0.7921 & 0.6151 & 0.8083 & 0.7964 & 0.8180 & \textbf{0.8197}  & 0.7357  \\
    & NDCG@10 $\uparrow$  & 0.5964 & 0.5912 & 0.5237 & 0.6071 & 0.5789 & \textbf{0.6293} & 0.6097 & 0.5984 \\
    & HitRate@20 $\uparrow$    & 0.8339 & 0.8268 & 0.6523 & 0.8344 & 0.8367 & 0.8374 & \textbf{0.8530}  & 0.8209  \\
    & NDCG@20 $\uparrow$  & 0.6146 & 0.6213 & 0.5354 & 0.6269 & 0.6091 & 0.6468 & \textbf{0.6554} & 0.6398 \\
    \bottomrule
    \bottomrule
  \end{tabular}}
\end{table*}

\subsubsection{Settings}

\textbf{Experimental Setup.} 
We conduct training and simulate incremental learning on public datasets as complementary experiments. The data is chronologically partitioned into training, validation, and test sets, with core data representing user-item interactions. For datasets lacking node features, we generate random features when necessary, construct heterogeneous relationships, and evaluate recall prediction on the test set.

In simulated incremental learning, we emulate online deployment through an initial training set and an incremental data stream. The initial set pretrains the model to convergence, while incremental batches are sequentially fed in fixed-size batches. As demonstrated in Sec. \ref{sec:workflow}, without full model retraining, each batch undergoes fine-tuning for model adaptation and a cold-start update for new node embeddings.

\noindent\textbf{Parameter Setting.} For experiments on public datasets, we use the model performance on the validation set for hyperparameter tuning across all models. Unless otherwise stated, the hyperparameters are selected via grid search, with the following search spaces:
\begin{itemize}
    \item Learning rate: \{5e-6, 1e-5, 5e-5, 1e-4, 5e-4\}
    \item Weight decay: \{1e-5, 5e-5, 1e-5, 5e-4, 1e-4\}
    \item Hidden size: \{16, 32, 64, 128, 256\}
    \item Dropout ratio: \{0.3, 0.4, 0.5, 0.6\}
    \item Number of layers: \{1, 2, 3, 4\}
    \item SGFormer Output Ratio: \{0.4, 0.5, 0.6, 0.8\}
    \item Limited Degree of Subgraph Sampling: \{5, 10, 20\}
\end{itemize}
For reproducibility, we also report one representative configuration. On the Ali-Display dataset, the final HetSGFormer model uses a learning rate of 5e-4, weight decay of 1e-4, a hidden size of 64, a dropout ratio of 0.5, three layers, an SGFormer output ratio of 0.6, and a limited degree of subgraph sampling as 10.

\noindent \textbf{Baselines.}
Given that ablation studies are detailed in Sec. \ref{sec:offline_experiment_setting}, our public benchmark comparison focuses on heterogeneous or recommendation-oriented encoders that can exploit node and edge types. To ensure a comprehensive evaluation and validate the performance of HetSGFormer+ILLE, we select a diverse set of representative baselines spanning three distinct paradigms, including traditional graph convolutional networks (e.g., LightGCN \cite{he2020lightgcn}, PinSAGE \cite{ying2018pinsage}), graph transformers (e.g., DiFFormer \cite{wu2023difformer}, GFormer \cite{li2023gformer}), and heterogeneous graph methods (e.g., HERec \cite{shi2018herec}, HGT \cite{hu2020hgt}), all described in detail in Sec. \ref{sec:related}. 

\noindent \textbf{Datasets.}
All models are evaluated on four public offline datasets, including Ali-Display, Epinions, Amazon-CD and Yelp. While these public benchmarks are relatively small, they serve to establish the feasibility of our approach and ensure comparability against widely used public baselines. The basic statistics of these datasets, along with their sparsities, are summarized in Tab. \ref{tab:dataset_stats}. The specific descriptions are provided below.

\begin{itemize}
    \item \textbf{Ali-Display}: User behaviors from Alibaba's display ads, with relationships derived from user and ad data.
    \item \textbf{Epinions}: User ratings for items in an online review system, with relationships based on user trust and item categories.
    \item \textbf{Amazon-CD}: User reviews for music on Amazon, with relationships from purchase history and item metadata.
    \item \textbf{Yelp}: Social connections, venue ratings, and merchant attributes for local business recommendations on Yelp.
\end{itemize}

\noindent \textbf{Computational Resource.}
The offline experiments were conducted on a computing workstation featuring a 16-core CPU, 64 GB of RAM, and an NVIDIA RTX 4090 GPU. The incremental training used an identical CPU/RAM configuration (16 cores, 64GB) but deliberately excluded GPU acceleration. This resource-constrained design simulates real-world environments where GPU scarcity weakens computational capacity during model updates.

\noindent \textbf{Metrics.}
In our evaluation settings, one positive (interacted) item and 99 negative (non-interacted) items are sampled for each user for performance evaluation. We use \textbf{HitRate@K} and \textbf{NDCG@K} as core metrics. HitRate@K is already introduced in Sec.\ref{sec:offline_experiment_setting}, and NDCG@K evaluates the quality of the ranking order by assigning higher scores to hits at upper ranks, reflecting the model's capability to prioritize highly relevant items at the most prominent positions, which is defined as:
\begin{equation}
  \mathrm{NDCG}@K = 
  \frac{1}{\lvert \mathcal{U}_{\mathrm{test}} \rvert} 
  \sum_{u \in \mathcal{U}_{\mathrm{test}}} 
  \frac{\sum_{i=1}^{K} \frac{\mathbb{I}[i_u = i]}{\log_2(i+1)}}{\sum_{i=1}^{\min(K, |\mathcal{I}_{u}|)} \frac{1}{\log_2(i+1)}},
\end{equation}
where $\mathcal{U}_{\mathrm{test}}$ is the set of test users, $i_u$ is the relevant items for user $u$, $\mathcal{R}^K_u$ is the top-$K$ recommendation list for user $u$, and $\mathcal{I}_{u}$ is the ground-truth item set for user $u$. In public dataset experiments, considering their relatively limited scale, we set K to 10 and 20.

\begin{table*}[!tb]
    \centering
    \caption{We compare leading scalable methods on industrial-scale graphs. A model's ability to jointly encode node attributes and identity is essential for cold-start scenarios where attributes are only partially available. We further assess whether the model can cope with sparse connections in both the static (SGL) and incremental (IGL) stages as a key indicator of its cold-start resilience. The running efficiency of the two stages is also compared.} 
    \resizebox{1\textwidth}{!}{
    \begin{tabular}{c|c|c|c|c|c|c|c}
    \toprule
        \multirow{2}{*}{Method} & \multicolumn{2}{c|}{Embedding Ability} & \multicolumn{2}{c|}{SGL stage} & \multicolumn{3}{c}{IGL stage} \\
        \cline{2-8}
        & Attributes & Identity & Sparse Connections & Training Complexity & Enabled & Sparse Connections & CPU-only Efficiency \\
        \midrule
        TransD~\cite{ji2015transd} & \XSolidBrush & \Checkmark & \XSolidBrush & $O(V+E)$ &  \multirow{3}{*}{\XSolidBrush} & \multirow{3}{*}{N/A} & \multirow{3}{*}{N/A} \\
        \cline{1-5} 
        PinSAGE~\cite{ying2018pinsage}  & \XSolidBrush & \Checkmark & \Checkmark (via random walk) & $O(V+E)$ & & \\
        \cline{1-5} 
        HGT~\cite{hu2020hgt} & \Checkmark & \XSolidBrush & \XSolidBrush & $O(V+E)$ & & \\
        \hline
        \makecell[c]{DyGFormer~\cite{yu2023towards}, \\DyHAN~\cite{yang2020dyhan}, DySAT~\cite{sankar2018dysat}} & \Checkmark & \XSolidBrush & \multicolumn{2}{c|}{N/A (no explicit SGL stage)} & \Checkmark & \Checkmark (via sequence modeling) & \makecell[c]{Low (computing attention \\ matrix is time-consuming)} \\
        \midrule
        \textbf{Ours} (Fig. \ref{fig:model}) & \Checkmark & \Checkmark & \Checkmark (via global attention) & $O(V+E)$ & \Checkmark & \Checkmark (via multi-hop sampling) & High (based on matrix factorization) \\
    \bottomrule
    \end{tabular}}
    \label{tab:comparison}
\end{table*}

\subsubsection{Results}

\textbf{Offline ranking performance.} As presented in Table \ref{tab:metrics_comparison_updated}, HetSGFormer demonstrates superior efficacy across the four diverse public datasets, consistently outperforming a broad spectrum of baselines ranging from traditional GNNs to advanced graph transformers. Specifically, HetSGFormer secures the top position in the vast majority of dataset-metric combinations. In the few instances where it ranks second, the performance margins are negligible, indicating high robustness. This consistent dominance can be attributed to our architecture's unique ability to synergize global and local information. Unlike standard GNN baselines, which are limited by the receptive field of local message passing, HetSGFormer leverages the GAL to capture long-range semantic dependencies, which are pivotal for sparse datasets. Conversely, compared to pure transformer-based models, our inclusion of local information ensures that fine-grained neighborhood topology is not lost amidst global aggregation. The results confirm that our hybrid design effectively mitigates the trade-off between capturing high-order graph structure and preserving local connectivity, yielding a generalized solution for heterogeneous graph recommendation.

\noindent  \textbf{Incremental adaptation efficiency.} The HetSGFormer + ILLE variant, which updates node representations incrementally without full retraining, yields slightly lower scores (typically within $5$--$10\%$ of the full HetSGFormer) but remains competitive with other baselines on all datasets. This slight drop is expected, as ILLE only touches a localized subset of parameters and avoids recomputing the entire model for each batch of new interactions. In return, ILLE enables rapid updates, keeping item and user representations up to date with low response latency. In practical recommendation scenarios where new items and behaviors continuously arrive, this trade-off can make updates more efficient, making our method a more realistic choice for deployment than purely offline retrained models.

\section{Related Works}\label{sec:related}

To better position our work, we briefly review three lines of related research: graph embedding, graph neural networks, and graph Transformers. We give a detailed technical comparison with important relevant methods in Table \ref{tab:comparison} as a summary.

\subsection{Graph Embedding} 
Graph embedding techniques map nodes and edges into low-dimensional vector spaces while preserving structural proximity, enabling a range of tasks such as node classification, link prediction, and knowledge graph completion. Early homogeneous methods, such as DeepWalk \cite{perozzi2014deepwalk} and Node2Vec \cite{grover2016node2vec}, utilize random walks and skip-gram models to capture topological context. To handle diverse entity types, heterogeneous methods such as HERec \cite{shi2018herec} leverage random walks over meta-paths to integrate multiple relationships. Similarly, in knowledge graph domains, translational distance models like TransD \cite{ji2015transd} introduce dynamic mapping matrices to capture complex relations. While these methods have demonstrated success across various tasks, they face significant challenges, including cold-start problems, sparsity, and scalability in real-world recommendation scenarios.

\subsection{Graph Neural Networks}
Graph Neural Networks have become popular in recommender systems due to the graph structures inherent in user-item interactions, with early methods using GCNs for user and item embeddings \cite{liu2020modelling, liu2022multi, gao2023survey}. LightGCN \cite{he2020lightgcn} simplifies traditional GCNs by removing non-linear activations and feature transformations, improving both efficiency and performance. GraphSAGE \cite{hamilton2017inductive} samples fixed-size neighborhoods for each node and uses an aggregator to update embeddings. Spatial models like PinSAGE \cite{ying2018pinsage}, which uses random walk-based sampling, are well-suited for large-scale applications. While these methods reduce computational complexity, they often lack representational power \cite{sharma2024survey}. Methods like SeRec \cite{chen2021efficient} attempt to handle heterogeneity by using specialized Heterogeneous GNN encoders. However, these approaches face challenges in generalizing across diverse interaction types and often lack adaptability to incorporate additional attribute edge types or hierarchical attributes. 

\subsection{Graph Transformer}
Transformers have revolutionized representation learning by employing self-attention mechanisms to capture long-range dependencies and enabling parallel training \cite{kreuzer2021rethinking, chen2022structure, hussain2022global}. The Graph Attention Network \cite{velivckovic2017graph} introduced attention in graphs. In graph learning, models like Graphormer \cite{lin2021mesh} and PolyFormer \cite{liu2023polyformer} integrate Transformers with static graph data, using positional and structural encodings for effective graph processing. 
\textbf{Dynamic Graph Transformers} including DySAT \cite{sankar2018dysat}, DyHAN \cite{yang2020dyhan}, and DyGFormer \cite{yu2023towards} focus on dynamic graph evolution. However, these models often lack interpretability, hindering progress. Models like DIFFormer \cite{wu2023difformer} apply dynamic interaction attention for sparse graphs, while GFormer \cite{li2023gformer} combines contrastive learning with transformer attention, and HGT \cite{hu2020hgt} designs type-aware attention for heterogeneous graphs. Despite their ability to capture global graph structures, Transformer models often struggle with high computational complexity and lack the ability to effectively handle large-scale heterogeneous data, posing significant challenges in practical applications.

\section{Conclusion}

Addressing the engineering challenges of billion-scale recommendation, we developed a two-stage dynamic graph embedding system. We tackled the SGL stage bottleneck with HetSGFormer, which ensures linear scalability and robust cold-start representations via attribute-ID fusion. For the IGL stage, we introduced ILLE, a novel CPU-only algorithm that enables rapid, incremental updates. Our experiments confirm the system's superiority: HetSGFormer delivered an up to 6.11\% lift in Advertiser Value compared to traditional baselines, while ILLE further enhanced HetSGFormer by 3.22\% and improved update timeliness by 83.2\%. Overall, this work provides a validated framework for deploying efficient dynamic graph learning in large-scale production environments.

\bibliographystyle{IEEEtran}
\bibliography{refer}

@article{wu2023sgformer,
  title={Sgformer: Simplifying and empowering transformers for large-graph representations},
  author={Wu, Qitian and Zhao, Wentao and Yang, Chenxiao and Zhang, Hengrui and Nie, Fan and Jiang, Haitian and Bian, Yatao and Yan, Junchi},
  journal={Advances in Neural Information Processing Systems},
  volume={36},
  pages={64753--64773},
  year={2023}
}

@inproceedings{ying2018pinsage,
  title={Graph convolutional neural networks for web-scale recommender systems},
  author={Ying, Rex and He, Ruining and Chen, Kaifeng and Eksombatchai, Pong and Hamilton, William L and Leskovec, Jure},
  booktitle={Proceedings of the 24th ACM SIGKDD international conference on knowledge discovery \& data mining},
  pages={974--983},
  year={2018}
}

@inproceedings{hu2020hgt,
  title={Heterogeneous graph transformer},
  author={Hu, Ziniu and Dong, Yuxiao and Wang, Kuansan and Sun, Yizhou},
  booktitle={Proceedings of the web conference 2020},
  pages={2704--2710},
  year={2020}
}

@inproceedings{ji2015transd,
  title={Knowledge graph embedding via dynamic mapping matrix},
  author={Ji, Guoliang and He, Shizhu and Xu, Liheng and Liu, Kang and Zhao, Jun},
  booktitle={Proceedings of the 53rd annual meeting of the association for computational linguistics and the 7th international joint conference on natural language processing (volume 1: Long papers)},
  pages={687--696},
  year={2015}
}

@article{wu2023difformer,
  title={DIFFormer: Scalable (Graph) Transformers Induced by Energy Constrained Diffusion},
  author={Wu, Qitian and Yang, Chenxiao and Zhao, Wentao and He, Yixuan and Wipf, David and Yan, Junchi},
  journal={The Eleventh International Conference on Learning Representations (ICLR)},
  year={2023}
}

@article{vaswani2017attention,
  title={Attention is all you need},
  author={Vaswani, Ashish and Shazeer, Noam and Parmar, Niki and Uszkoreit, Jakob and Jones, Llion and Gomez, Aidan N and Kaiser, {\L}ukasz and Polosukhin, Illia},
  journal={Advances in neural information processing systems},
  volume={30},
  year={2017}
}

@inproceedings{he2020lightgcn,
  title={Lightgcn: Simplifying and powering graph convolution network for recommendation},
  author={He, Xiangnan and Deng, Kuan and Wang, Xiang and Li, Yan and Zhang, Yongdong and Wang, Meng},
  booktitle={Proceedings of the 43rd International ACM SIGIR conference on research and development in Information Retrieval},
  pages={639--648},
  year={2020}
}

@inproceedings{he2017neural,
  title={Neural collaborative filtering},
  author={He, Xiangnan and Liao, Lizi and Zhang, Hanwang and Nie, Liqiang and Hu, Xia and Chua, Tat-Seng},
  booktitle={Proceedings of the 26th international conference on world wide web},
  pages={173--182},
  year={2017}
}

@article{wu2021representing,
  title={Representing long-range context for graph neural networks with global attention},
  author={Wu, Zhanghao and Jain, Paras and Wright, Matthew and Mirhoseini, Azalia and Gonzalez, Joseph E and Stoica, Ion},
  journal={Advances in neural information processing systems},
  volume={34},
  pages={13266--13279},
  year={2021}
}

@article{johnson2017fbann,
  title={Billion-scale similarity search with GPUs},
  author={Johnson, Jeff and Douze, Matthijs and J{\'e}gou, Herv{\'e}},
  journal={IEEE Transactions on Big Data},
  volume={7},
  number={3},
  pages={535--547},
  year={2019},
  publisher={IEEE}
}

@article{bastos2022softmax,
  title={How Expressive are Transformers in Spectral Domain for Graphs?},
  author={Bastos, Anson and Nadgeri, Abhishek and Singh, Kuldeep and Kanezashi, Hiroki and Suzumura, Toyotaro and Mulang, Isaiah Onando},
  journal={Transactions on Machine Learning Research},
  year={2022}
}

@article{shi2018herec,
  title={Heterogeneous information network embedding for recommendation},
  author={Shi, Chuan and Hu, Binbin and Zhao, Wayne Xin and Yu, Philip S},
  journal={IEEE transactions on knowledge and data engineering},
  volume={31},
  number={2},
  pages={357--370},
  year={2018},
  publisher={IEEE}
}

@inproceedings{li2023gformer,
  title={Graph transformer for recommendation},
  author={Li, Chaoliu and Xia, Lianghao and Ren, Xubin and Ye, Yaowen and Xu, Yong and Huang, Chao},
  booktitle={Proceedings of the 46th international ACM SIGIR conference on research and development in information retrieval},
  pages={1680--1689},
  year={2023}
}

@article{kipf2016semigcn,
  title={Semi-Supervised Classification with Graph Convolutional Networks},
  author={Kipf, TN},
  journal={International Conference on Learning Representations (ICLR)},
  year={2017}
}

@article{liu2022multi,
  title={Multi-perspective social recommendation method with graph representation learning},
  author={Liu, Hai and Zheng, Chao and Li, Duantengchuan and Zhang, Zhaoli and Lin, Ke and Shen, Xiaoxuan and Xiong, Neal N and Wang, Jiazhang},
  journal={Neurocomputing},
  volume={468},
  pages={469--481},
  year={2022},
  publisher={Elsevier}
}

@article{liu2020modelling,
  title={Modelling high-order social relations for item recommendation},
  author={Liu, Yang and Chen, Liang and He, Xiangnan and Peng, Jiaying and Zheng, Zibin and Tang, Jie},
  journal={IEEE Transactions on Knowledge and Data Engineering},
  volume={34},
  number={9},
  pages={4385--4397},
  year={2020},
  publisher={IEEE}
}

@article{hamilton2017inductive,
  title={Inductive representation learning on large graphs},
  author={Hamilton, Will and Ying, Zhitao and Leskovec, Jure},
  journal={Advances in neural information processing systems},
  volume={30},
  year={2017}
}

@article{sharma2024survey,
  title={A survey of graph neural networks for social recommender systems},
  author={Sharma, Kartik and Lee, Yeon-Chang and Nambi, Sivagami and Salian, Aditya and Shah, Shlok and Kim, Sang-Wook and Kumar, Srijan},
  journal={ACM Computing Surveys},
  volume={56},
  number={10},
  pages={1--34},
  year={2024},
  publisher={ACM New York, NY}
}

@inproceedings{chen2021efficient,
  title={An efficient and effective framework for session-based social recommendation},
  author={Chen, Tianwen and Wong, Raymond Chi-Wing},
  booktitle={Proceedings of the 14th ACM international conference on web search and data mining},
  pages={400--408},
  year={2021}
}

@inproceedings{perozzi2014deepwalk,
  title={Deepwalk: Online learning of social representations},
  author={Perozzi, Bryan and Al-Rfou, Rami and Skiena, Steven},
  booktitle={Proceedings of the 20th ACM SIGKDD international conference on Knowledge discovery and data mining},
  pages={701--710},
  year={2014}
}

@inproceedings{grover2016node2vec,
  title={node2vec: Scalable feature learning for networks},
  author={Grover, Aditya and Leskovec, Jure},
  booktitle={Proceedings of the 22nd ACM SIGKDD international conference on Knowledge discovery and data mining},
  pages={855--864},
  year={2016}
}

@inproceedings{lin2021mesh,
  title={Mesh graphormer},
  author={Lin, Kevin and Wang, Lijuan and Liu, Zicheng},
  booktitle={Proceedings of the IEEE/CVF international conference on computer vision},
  pages={12939--12948},
  year={2021}
}

@inproceedings{liu2023polyformer,
  title={Polyformer: Referring image segmentation as sequential polygon generation},
  author={Liu, Jiang and Ding, Hui and Cai, Zhaowei and Zhang, Yuting and Satzoda, Ravi Kumar and Mahadevan, Vijay and Manmatha, R},
  booktitle={Proceedings of the IEEE/CVF conference on computer vision and pattern recognition},
  pages={18653--18663},
  year={2023}
}

@article{velivckovic2017graph,
  title={Graph attention networks},
  author={Veli{\v{c}}kovi{\'c}, Petar and Cucurull, Guillem and Casanova, Arantxa and Romero, Adriana and Lio, Pietro and Bengio, Yoshua},
  journal={International Conference on Learning Representations (ICLR)},
  year={2018}
}

@article{yu2023towards,
  title={Towards better dynamic graph learning: New architecture and unified library},
  author={Yu, Le and Sun, Leilei and Du, Bowen and Lv, Weifeng},
  journal={Advances in Neural Information Processing Systems},
  volume={36},
  pages={67686--67700},
  year={2023}
}

@article{kreuzer2021rethinking,
  title={Rethinking graph transformers with spectral attention},
  author={Kreuzer, Devin and Beaini, Dominique and Hamilton, Will and L{\'e}tourneau, Vincent and Tossou, Prudencio},
  journal={Advances in Neural Information Processing Systems},
  volume={34},
  pages={21618--21629},
  year={2021}
}

@article{sankar2018dysat,
  title={Dynamic graph representation learning via self-attention networks},
  author={Sankar, Aravind and Wu, Yanhong and Gou, Liang and Zhang, Wei and Yang, Hao},
  journal={Workshop on Representation Learning on Graphs and Manifolds, International Conference on Learning Representations (ICLR)},
  year={2019}
}

@inproceedings{yang2020dyhan,
  title={Dynamic heterogeneous graph embedding using hierarchical attentions},
  author={Yang, Luwei and Xiao, Zhibo and Jiang, Wen and Wei, Yi and Hu, Yi and Wang, Hao},
  booktitle={European conference on information retrieval},
  pages={425--432},
  year={2020},
  organization={Springer}
}

@inproceedings{chen2022structure,
  title={Structure-aware transformer for graph representation learning},
  author={Chen, Dexiong and O’Bray, Leslie and Borgwardt, Karsten},
  booktitle={International conference on machine learning},
  pages={3469--3489},
  year={2022},
  organization={PMLR}
}

@article{wang2022survey,
  title={A survey on heterogeneous graph embedding: methods, techniques, applications and sources},
  author={Wang, Xiao and Bo, Deyu and Shi, Chuan and Fan, Shaohua and Ye, Yanfang and Yu, Philip S},
  journal={IEEE transactions on big data},
  volume={9},
  number={2},
  pages={415--436},
  year={2022},
  publisher={IEEE}
}

@inproceedings{ji2021dynamic,
  title={Dynamic heterogeneous graph embedding via heterogeneous hawkes process},
  author={Ji, Yugang and Jia, Tianrui and Fang, Yuan and Shi, Chuan},
  booktitle={Joint European Conference on Machine Learning and Knowledge Discovery in Databases},
  pages={388--403},
  year={2021},
  organization={Springer}
}

@article{deng2022recommender,
  title={Recommender systems based on graph embedding techniques: A review},
  author={Deng, Yue},
  journal={IEEE Access},
  volume={10},
  pages={51587--51633},
  year={2022},
  publisher={IEEE}
}

@article{sha2021commercial,
  title={Hierarchical attentive knowledge graph embedding for personalized recommendation},
  author={Sha, Xiao and Sun, Zhu and Zhang, Jie},
  journal={Electronic Commerce Research and Applications},
  volume={48},
  pages={101071},
  year={2021},
  publisher={Elsevier}
}

@inproceedings{zhang2013optimizing,
  title={Optimizing top-n collaborative filtering via dynamic negative item sampling},
  author={Zhang, Weinan and Chen, Tianqi and Wang, Jun and Yu, Yong},
  booktitle={Proceedings of the 36th international ACM SIGIR conference on Research and development in information retrieval},
  pages={785--788},
  year={2013}
}

@article{roweis2000nonlinear,
  title={Nonlinear dimensionality reduction by locally linear embedding},
  author={Roweis, Sam T and Saul, Lawrence K},
  journal={science},
  volume={290},
  number={5500},
  pages={2323--2326},
  year={2000},
  publisher={American Association for the Advancement of Science}
}

@article{kouropteva2005incremental,
  title={Incremental locally linear embedding},
  author={Kouropteva, Olga and Okun, Oleg and Pietik{\"a}inen, Matti},
  journal={Pattern recognition},
  volume={38},
  number={10},
  pages={1764--1767},
  year={2005},
  publisher={Elsevier}
}

@inproceedings{higgins2017beta,
  title={beta-vae: Learning basic visual concepts with a constrained variational framework},
  author={Higgins, Irina and Matthey, Loic and Pal, Arka and Burgess, Christopher and Glorot, Xavier and Botvinick, Matthew and Mohamed, Shakir and Lerchner, Alexander},
  booktitle={International conference on learning representations},
  year={2017}
}

@article{chen2011locally,
  title={Locally linear embedding: a survey},
  author={Chen, Jing and Liu, Yang},
  journal={Artificial Intelligence Review},
  volume={36},
  number={1},
  pages={29--48},
  year={2011},
  publisher={Springer}
}

@article{chang2006robust,
  title={Robust locally linear embedding},
  author={Chang, Hong and Yeung, Dit-Yan},
  journal={Pattern recognition},
  volume={39},
  number={6},
  pages={1053--1065},
  year={2006},
  publisher={Elsevier}
}

@article{gao2023survey,
  title={A survey of graph neural networks for recommender systems: Challenges, methods, and directions},
  author={Gao, Chen and Zheng, Yu and Li, Nian and Li, Yinfeng and Qin, Yingrong and Piao, Jinghua and Quan, Yuhan and Chang, Jianxin and Jin, Depeng and He, Xiangnan and others},
  journal={ACM Transactions on Recommender Systems},
  volume={1},
  number={1},
  pages={1--51},
  year={2023},
  publisher={ACM New York, NY, USA}
}

@article{hou2009local,
  title={Local linear transformation embedding},
  author={Hou, Chenping and Wang, Jing and Wu, Yi and Yi, Dongyun},
  journal={Neurocomputing},
  volume={72},
  number={10-12},
  pages={2368--2378},
  year={2009},
  publisher={Elsevier}
}

@article{arya1998optimal,
  title={An optimal algorithm for approximate nearest neighbor searching fixed dimensions},
  author={Arya, Sunil and Mount, David M and Netanyahu, Nathan S and Silverman, Ruth and Wu, Angela Y},
  journal={Journal of the ACM (JACM)},
  volume={45},
  number={6},
  pages={891--923},
  year={1998},
  publisher={ACM New York, NY, USA}
}

@inproceedings{hussain2022global,
  title={Global self-attention as a replacement for graph convolution},
  author={Hussain, Md Shamim and Zaki, Mohammed J and Subramanian, Dharmashankar},
  booktitle={Proceedings of the 28th ACM SIGKDD conference on knowledge discovery and data mining},
  pages={655--665},
  year={2022}
}

@inproceedings{grad2017graph,
  title={Graph embedding based recommendation techniques on the knowledge graph},
  author={Grad-Gyenge, L{\'a}szl{\'o} and Kiss, Attila and Filzmoser, Peter},
  booktitle={Adjunct publication of the 25th conference on user modeling, adaptation and personalization},
  pages={354--359},
  year={2017}
}
\begin{IEEEbiography}[{\includegraphics[width=1in,height=1.25in,clip,keepaspectratio]{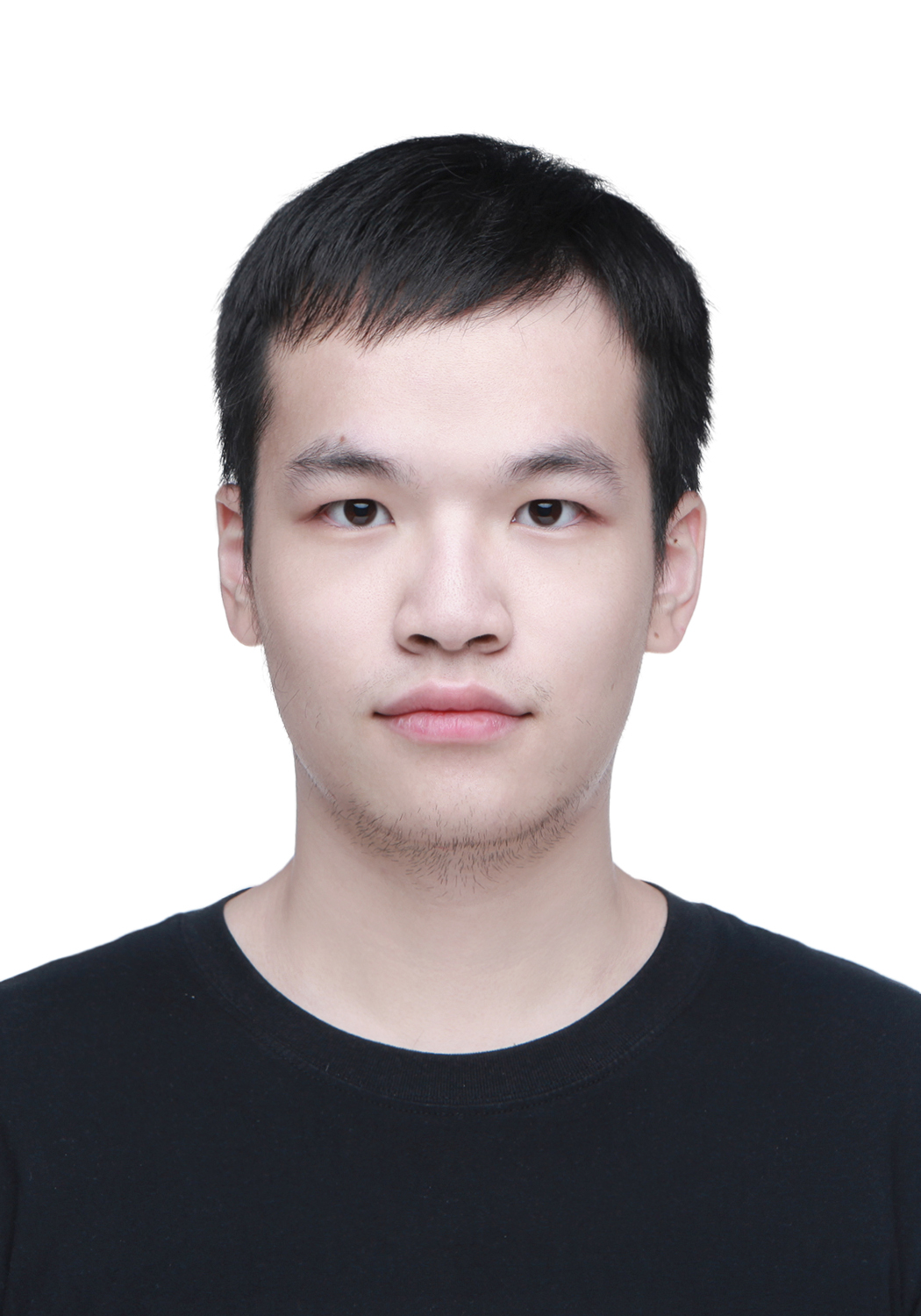}}]{Mabiao Long} is currently a graduate student with Department of Computer Science and Engineering, Shanghai Jiao Tong University, under the supervision of Prof. Junchi Yan. He obtained his B.E. degree in Computer Science from the same university in 2024. His research interests include graph machine learning and quantum machine learning in AI4Science.
\end{IEEEbiography}

\begin{IEEEbiography}[{\includegraphics[width=1in,height=1.25in,clip,keepaspectratio]{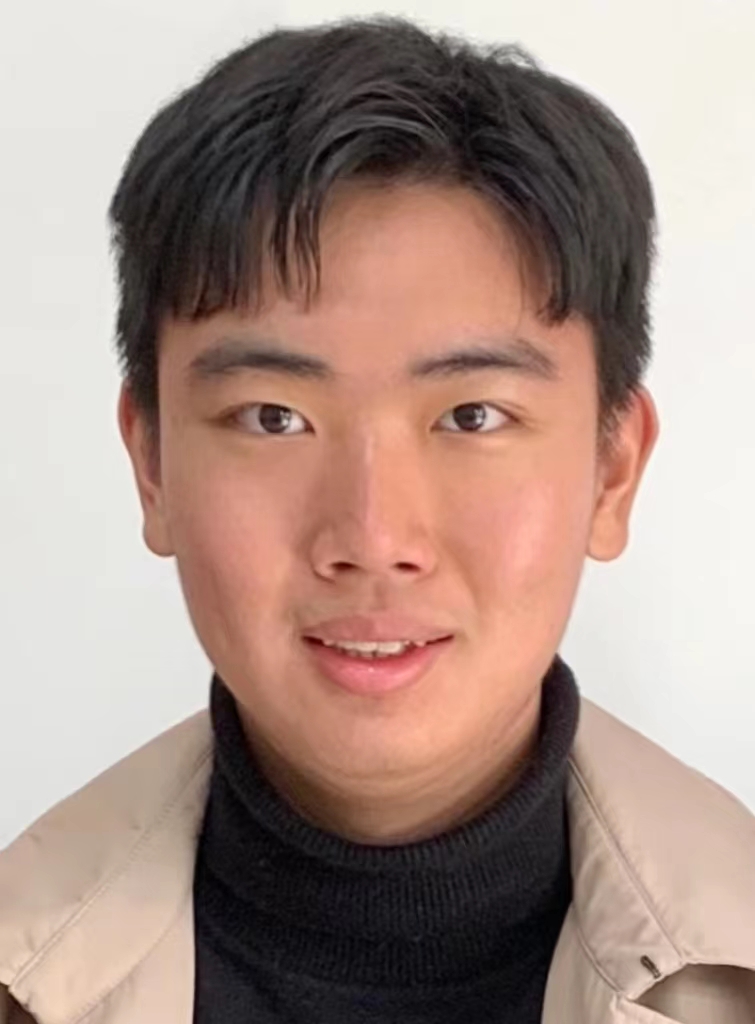}}]{Jiaxi Liu} is currently a first-year graduate student with Department of Computer Science and Engineering, Shanghai Jiao Tong University, under the supervision of Prof. Junchi Yan. Before that, he received the B.E. degree in Artificial Intelligence from Nanjing University in 2025. His research interests concentrate on graph machine learning and large language model applications.
\end{IEEEbiography}

\begin{IEEEbiography}[{\includegraphics[width=1in,height=1.25in,clip,keepaspectratio]{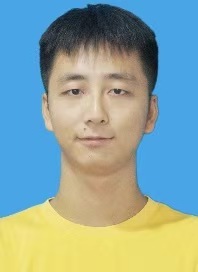}}]{Yufeng Li} is currently a Ph.D. candidate in the Department of Computer Science and Engineering at Shanghai Jiao Tong University, under the supervision of Prof. Junchi Yan. His research interests include graph machine learning, embodied intelligence, and optimal transport theory.
\end{IEEEbiography}

\begin{IEEEbiography}[{\includegraphics[width=1in,height=1.25in,clip,keepaspectratio]{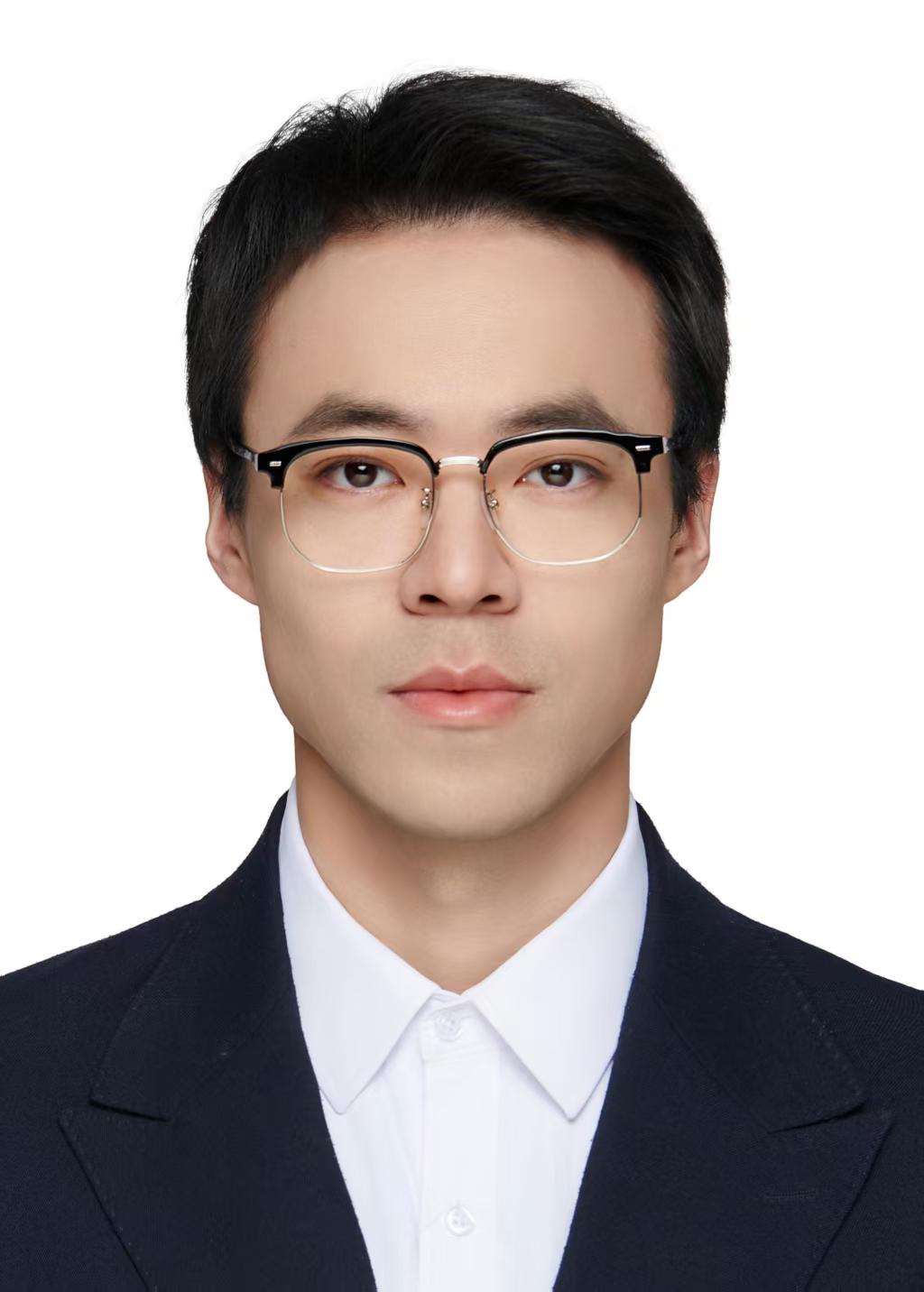}}]{Hao Xiong} is currently an Assistant Professor at AI$^3$ (Artificial Intelligence Innovation and Incubation) Institute of Fudan University, Shanghai, China. Before that, he received the Ph.D. degree in Computer Science from Shanghai Jiao Tong University in 2025, under the supervision of Prof. Junchi Yan. His research interests include graph machine learning and quantum artificial intelligence.
\end{IEEEbiography}

\begin{IEEEbiography}[{\includegraphics[width=1in,height=1.25in,clip,keepaspectratio]{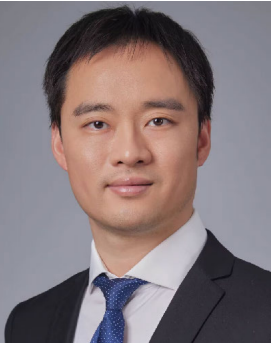}}]{Junchi Yan} (S'10-M'11-SM'21) is currently a Full Professor with AI Institute, Shanghai Jiao Tong University. Before that, he was a Senior Research Staff Member and Principal Scientist with IBM Research -- China where he started his career in April 2011. He obtained the Ph.D. in Electrical Engineering, from Shanghai Jiao Tong University, China in 2015. His research interests are machine learning and computer vision. He serves as  Area Chair for ACM-MM 2021/22, CVPR 2021, AAAI 2022, ICML 2022. He is a Senior Member of IEEE.
\end{IEEEbiography}

\begin{IEEEbiography}[{\includegraphics[width=1in,height=1.25in,clip,keepaspectratio]{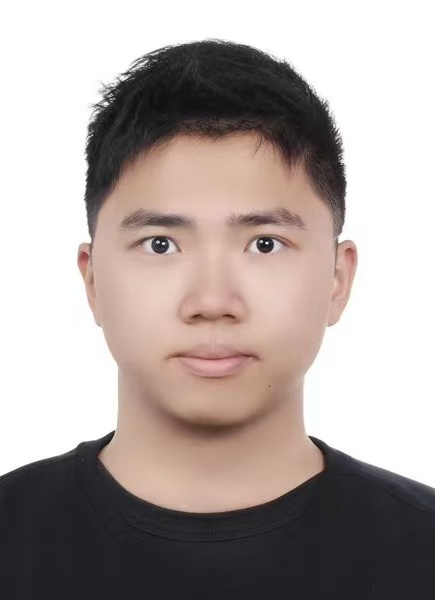}}]{Kefan Wang} is currently an independent researcher. He received his M.Sc. degree in Data Science and Machine Learning from University College London in 2022 and his B.Sc. degree in Computer Science from the same university in 2021, both under the supervision of Dr. Denise Gorse. His research interests include machine learning in recommendation systems and web search algorithms.
\end{IEEEbiography}

\begin{IEEEbiography}[{\includegraphics[width=1in,height=1.25in,clip,keepaspectratio]{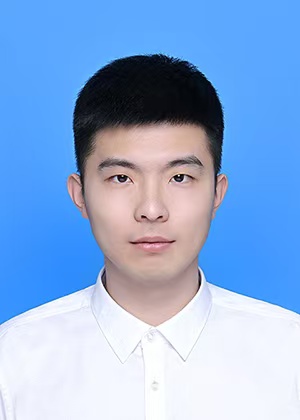}}]{Yi Cao} is currently an independent researcher. He received his Ph.D degree in Information and Communication Engineering from Xidian University in 2020. His research interests include recommendation systems, search algorithms and pattern recognition.
\end{IEEEbiography}

\begin{IEEEbiography}[{\includegraphics[width=1in,height=1.25in,clip,keepaspectratio]{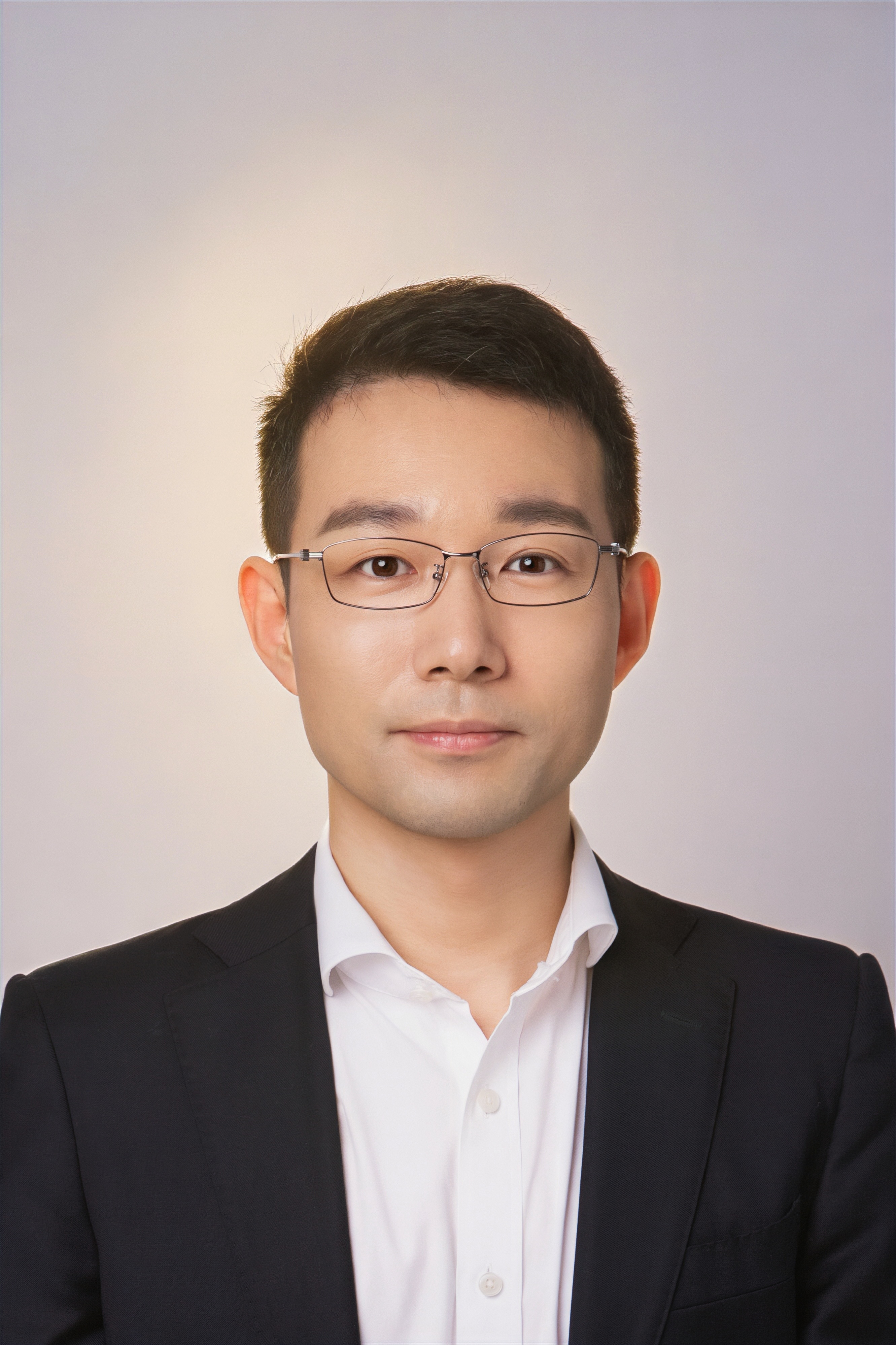}}]{Jiandong Ding} is an Enterprise Mentor at Fudan University, Shanghai. He received his Ph.D. degree in Computer Science from the Department of Computer Science at Fudan University in 2012. His research interests include recommender systems, data mining, and AI-driven decision-making. He has published 18 papers in top-tier journals and conferences and holds 22 granted patents in the fields of AI and data engineering.
\end{IEEEbiography}

\end{document}